\documentclass[final,5p,times,twocolumn]{elsarticle}

\usepackage{graphicx}
\usepackage{amssymb}

\usepackage{lineno}

\usepackage{amsmath}

\usepackage{makecell} 

\usepackage{floatrow} 
\newfloatcommand{capbtabbox}{table}[][\FBwidth]

\usepackage{capt-of}

\usepackage[colorinlistoftodos]{todonotes}  

\usepackage{url}
\usepackage{hyperref}

\usepackage{tabularx}

\usepackage{float} 
\usepackage[caption = false]{subfig}

\newcommand\footnoteref[1]{\protected@xdef\@thefnmark{\ref{#1}}\@footnotemark}

\usepackage{booktabs} 

\usepackage{rotating} 

\usepackage{framed}
\usepackage{nomencl} 
\makenomenclature

\newcommand{\nomunit}[1]{%
\renewcommand{\nomentryend}{\hspace*{\fill}#1}}

\usepackage{etoolbox}
\renewcommand\nomgroup[1]{%
  \item[\bfseries
  \ifstrequal{#1}{A}{Selected abbreviations}{%
  \ifstrequal{#1}{V}{Selected Variables, Parameters and Indices}{%
  \ifstrequal{#1}{P}{Parameters}{}}}%
]}

\biboptions{sort}

\usepackage{array}
\usepackage{multirow}
\usepackage{nicefrac}

\usepackage{amsmath,booktabs,dcolumn,caption}
\newcolumntype{d}[1]{D{.}{.}{#1}}

\usepackage{siunitx}
\usepackage{eurosym}

\DeclareSIUnit{\pers}{pers}
\DeclareSIUnit{\EUR}{\text{\euro}}
\usepackage[utf8x]{inputenc}
\usepackage{pgf} 

    \makeatletter
    \def\ps@pprintTitle{%
       \let\@oddhead\@empty
       \let\@evenhead\@empty
       \def\@oddfoot{\reset@font\hfil\thepage\hfil}
       \let\@evenfoot\@oddfoot
    }
    \makeatother

\usepackage{changes}
\definechangesauthor[name={Wilko}, color=blue]{WH}
\definechangesauthor[name={Wided}, color=red]{WM}

\begin{document}

\begin{frontmatter}


\title{Assessment of the regionalised demand response potential in Germany using an open source tool and dataset}



\author[label1]{Wilko Heitkoetter}\author[label1]{Bruno U. Schyska}\author[label1]{Danielle Schmidt}\author[label1]{Wided Medjroubi}\author[label1]{Thomas Vogt}\author[label1]{Carsten Agert} 
\address[label1]{DLR Institute of Networked Energy Systems, Carl-von-Ossietzky-Str. 15, Oldenburg, Germany, wilko.heitkoetter@dlr.de}

\begin{abstract}
With the expansion of renewable energies in Germany, imminent grid congestion events occur more often. One approach for avoiding curtailment of renewable energies is to cover excess feed-in by demand response.
As curtailment is often a local phenomenon, in this work we determine the regional demand response potential 
for the 401 German administrative districts.
The load regionalisation is based on weighting factors derived from population and employment statistics, locations of industrial facilities, etc.
Using periodic and temperature-dependent load profiles and technology specific parameters, e.g., the time frame of management, load shifting potentials were determined with a temporal resolution of 15 minutes.   
Our analysis yields that power-to-heat technologies provide the highest potentials, followed by residential appliances, commercial and industrial loads. For the considered 2030 scenario, power-to-gas and e-mobility also contribute a significant potential. 
The cumulated load increase potential of all technologies ranges from $5 - 470~MW$ per administrative district.
The median value is $25~MW$, which would suffice to avoid the curtailment of 8 classical wind turbines.
Further, we calculated load shifting cost-potential curves for each district. 
Industrial processes and power-to-heat in district heating have the lowest load shifting investment cost, due to the largest installed capacities per facility. 
We distinguished between different size classes of the installed capacity of heat pumps, yielding $23\%$ lower average investment cost for heat pump flexibilisation in the city of Berlin compared to a rural district. 
The variable costs of most considered load shifting technologies remain under the average compensation costs for curtailment of renewable energies of $110~\text{\euro{}}/MWh$.   
As all results and the developed code are published under open source licenses, they can be integrated into energy system models, which is simplified by the applied storage equivalent demand response formulation.
\end{abstract}


\end{frontmatter}

\section{Introduction}

In pursuit of reducing global CO$_2$ emissions and mitigating climate change, renewable energy sources are considered a main instrument~\cite{sims2004renewable}, constituting 33\% of global generation capacity in 2018~\citep{REN21}. However, the intermittent and non-dispatchable feed-in of variable renewable energy sources (VRE)~\cite{luz2019100} requires balancing technologies, such as dispatchable generators, energy storage or transmission line reinforcement~\cite{mueller2018demand}.

Another balancing option that plays a minor role currently~\cite{valdes2019industry}, but may gain importance with increasing shares of renewable energies, is demand response (DR)~\cite{moura2010role}. 
Demand response utilises available elasticity of consumer demand and comprises two classes~\cite{gils2015balancing}: Load shedding applies to loads being reduced, but for which cannot be compensated for at another time~\cite{paulus2011potential}. 
Load shifting is associated with loads being shifted to an earlier or later time, e.g., from a period with low VRE feed-in to a period with high VRE feed-in~\cite{mueller2018demand}.    
In this paper we focus on the application of demand response for avoiding VRE curtailment in times where feed-in exceeds demand or grid capacity. As this service can only be provided by load shifting, load shedding is disregarded in our analysis.

Table~\ref{tab_literature_overview} shows selected studies that provide spatially and temporally resolved data for the load shifting potential in Germany. 
The studies~\cite{stadler2005demand, klobasa2007dynamisch, paulus2011potential, ewi2012untersuchungen, apel2012notwendiger, pellinger2016merit, mueller2018demand, steurer2017analyse} analyse the potential for Germany, while~\cite{soder2018review} focuses on multiple countries in Northern Europe and the geographical scope of~\cite{gils2015balancing, vogt2016restore} is entire Europe.
Most of the mentioned studies determine aggregated load shifting potential values on country level (NUTS-0~\cite{destatis2020nuts}). Such a spatial resolution is sufficient, if the influence of load shifting on national power markets shall be investigated. For example, \citet{klobasa2007dynamisch} addresses decreased planning horizons in conventional power plant dispatch due to increased VRE penetration and assesses how the overall power system efficiency can be increased by the application of load shifting.

\begin{table*}[h!]
\caption{Overview of selected studies taking into account the spatially and temporally resolved load shifting potential in Germany.}
\scriptsize
\begin{center}
\begin{tabularx}{\textwidth}{X| X X X X X}
\hline
Study & 
Spatial scope/\newline spatial resolution/\newline temp. resolution& Electricity sector:\newline residential/ \newline commercial/ \newline industrial & Sector coupling:\newline power-to-heat/\newline e-mobility/\newline power-to-gas &
Potential restrictions:\newline technical/\newline socio-technical/\newline economic &
Openness and\newline reproducibility\\
\hline
\citet{stadler2005demand}& GER/NUTS-0/h  & \checkmark/\checkmark/\checkmark  & \checkmark/\textendash/\textendash & \checkmark/(\checkmark)/(\checkmark) & grey box\\
\citet{klobasa2007dynamisch}& GER/NUTS-0/\textendash  & \checkmark/\checkmark/\checkmark  & \checkmark/\textendash/\textendash & \checkmark/(\checkmark)/\checkmark & grey box\\
Paulus et al.~\cite{paulus2011potential}& GER/NUTS-0/\textendash  & \textendash/\textendash/\checkmark  & \checkmark/\textendash/\textendash & \checkmark/\textendash/\checkmark & grey box\\
\citet{ewi2012untersuchungen}& GER/NUTS-0/h  & \checkmark/\checkmark/\checkmark  & \checkmark/\checkmark/\textendash & \checkmark/(\checkmark)/\checkmark & grey box\\
\citet{apel2012notwendiger}& GER/NUTS-0/\textendash  & \checkmark/\checkmark/\checkmark & \checkmark/\textendash/\textendash & \checkmark/(\checkmark)/(\checkmark) & grey box\\
\citet{gils2015balancing}& EUR/NUTS-3/h  & \checkmark/\checkmark/\checkmark & \checkmark/\checkmark/\textendash & \checkmark/\textendash/\checkmark & grey box\\
\citet{vogt2016restore}& EUR/NUTS-0/h & \checkmark/\checkmark/\checkmark  & \checkmark/\checkmark/\textendash & \checkmark/\textendash/\textendash & grey box\\
\citet{soder2018review}& N.EUR/NUTS-0/\textendash  & \checkmark/\checkmark/\checkmark  & \checkmark/\textendash/\textendash & \checkmark/\textendash/\textendash & grey box\\
\citet{pellinger2016merit}& GER/NUTS-3/\textendash  & \checkmark/\checkmark/\checkmark  & \checkmark/\checkmark/\checkmark & \checkmark/(\checkmark)/\checkmark & grey box\\
M{\"u}ller et al.~\cite{mueller2018demand}& GER/NUTS-0/h  & \checkmark/\checkmark/\checkmark  & \checkmark/\textendash/\textendash & \checkmark/\textendash/\textendash & grey box\\
\citet{steurer2017analyse}& GER/NUTS-1/h  & \checkmark/\checkmark/\checkmark  & \checkmark/\textendash/\textendash & \checkmark/\checkmark/\checkmark & grey box\\
Present study& GER/NUTS-3/15min  & \checkmark/\checkmark/\checkmark  & \checkmark/\checkmark/\checkmark & \checkmark/\checkmark/\checkmark & white box\\
\hline
\multicolumn{6}{l}{Annotation: - = not considered; \checkmark = considered; (\checkmark) = limited consideration;}\\
\multicolumn{6}{l}{white box = applied equations, source code and data are publicly available; grey box = significant part of applied equations, source code or data remains undisclosed}
\end{tabularx}%
\label{tab_literature_overview}
\end{center}
\end{table*}

In the case that load shifting for the avoidance of VRE curtailment shall be assessed, a higher spatial resolution is required. VRE curtailment is mostly caused by surplus feed-in and limited grid transport capacity. As these factors depend on the installed capacity and energy demand in a specific region, as well as on the power flow from or to other regions~\cite{nep2017nep}, VRE curtailment is often a local phenomenon. The studies~\cite{gils2015balancing, pellinger2016merit} account for such a higher spatial resolution and provide load shifting potential values on administrative district level (NUTS-3~\cite{destatis2020nuts}). We henceforth denote the process of allocating data to regions of a territory as ``regionalisation".

The selected studies also differ with respect to the temporal resolution of the results. In~\cite{klobasa2007dynamisch,paulus2011potential,apel2012notwendiger,soder2018review,pellinger2016merit} no temporal resolution of the data is taken into account and instead minimum and maximum values for the load shifting potential are provided. The resulting potential data in \cite{stadler2008gigantisches, ewi2012untersuchungen, steurer2017analyse} are given with a temporal resolution of 1 hour.

Processes that are suitable for demand response typically provide thermal inertia, a physical storage or demand flexibility~\cite{gils2015balancing}. 
In the residential sector, white goods such as washing and drying machines, or fridges and freezers, are considered as suitable. Common DR applications in the commercial sector are ventilation and cooling appliances. In the industrial sector there are processes with a physical storage that are particularly fitted for load shifting, e.g., cement mills or wood pulpers. Cross-sectional technologies are associated with multiple industry branches, e.g., compressors for pressurised air. Load shedding processes regularly run at their full installed capacity and can thus only be switched off, not shifted, e.g. metal production. Most of the considered studies take into account all of the DR application categories, except~\cite{paulus2011potential}, which only focuses on the industrial sector.

In the course of decarbonisation the power sector will increasingly be coupled with the heating, transport and gas sector~\cite{victoria2019role}. While power-to-heat (PtH) technologies are considered in all investigated studies, the e-mobility sector is included on NUTS-0 level only by \citet{ewi2012untersuchungen} and spatially resolved results on NUTS-3 level are only provided by~\cite{gils2015balancing} and~\cite{pellinger2016merit}. The power-to-gas sector is only considered by~\citet{pellinger2016merit}.

The theoretical load shifting potential is further limited by technical restrictions, e.g. maximum load increase and decrease factors~\cite{steurer2017analyse}. Considering additional acceptance and organisational constraints yields the socio-technical potential. The economic potential is calculated by taking cost and revenue parameters into account.
While all analysed studies regard technical restrictions, some of the studies do not consider socio-technical and economic restrictions, or only consider these in a limited manner, e.g., by not providing explicit restriction factors. 

In the present study all mentioned technologies and potential restrictions are considered. Additionally, we include centralised PtH in district heating and in the industry, which was not regarded in the considered studies. For modelling residential PtH more than 700 building types are used to account for the regional differences in the residential building stock. The load shifting potential results are provided with a high spatial resolution on NUTS-3 level and a temporal resolution of 15 minutes. We determine the economic potential in the form of regional cost-potential curves for each administrative district. As a case study, we investigate the influence of heat pump size classes on the regional cost-potential curves. In contrast to other studies, all result data and developed source code are published open source.\footnote{\label{fn_sup_mat}For data and source code refer to the supplementary material of this article at: \href{https://doi.org/10.5281/zenodo.3988921}{https://doi.org/10.5281/zenodo.3988921}} This supports external model evaluation by other researchers and avoids duplication in data collection and code implementation~\cite{medjroubi2017open}. In particular, this paper will examine the following research questions:
\begin{itemize}
\item What are the load shifting potentials on NUTS-3 level in Germany?
\item How does the potential differ between rural and urban districts?
\item How are the regional load shifting cost-potential curves characterised?
\item How will the load shifting potential develop in future? 
\end{itemize}

\begin{figure*}[h!]
\centering\includegraphics[width=0.8\linewidth]{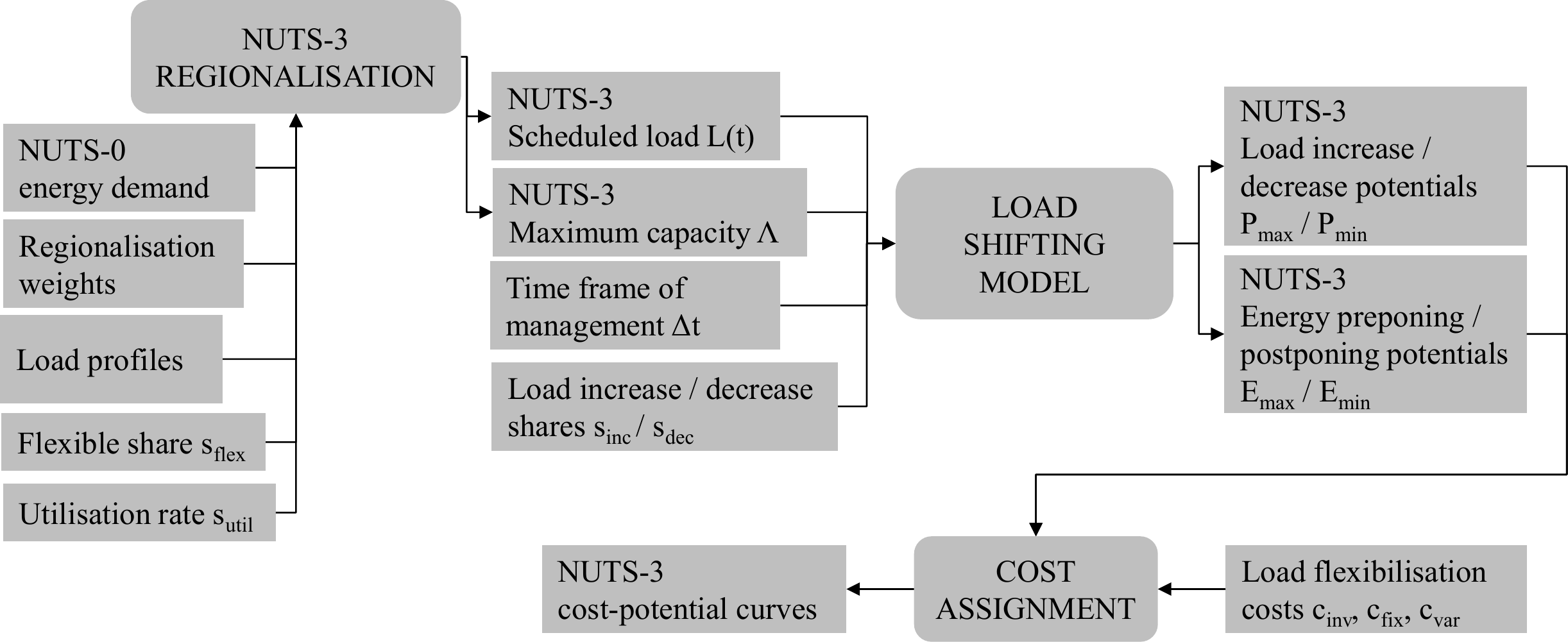}
\caption{Overview of the process for deriving the regionalised load shifting potentials and cost-potential cuvces, implemented in the open source tool \href{https://doi.org/10.5281/zenodo.3988921}{dsmlib}.}
\label{fig_flow_chart}
\end{figure*}
\section{Methods}\label{sec_methods}
The methodology of this work comprises the NUTS-3 load regionalisation and the determination of the load shifting potentials, as depicted in Figure~\ref{fig_flow_chart}. We implemented the computation in the open source python tool dsmlib, which can be obtained from the supplementary material, together with the used input data and obtained result data\textsuperscript{\ref{fn_sup_mat}}. 

In this section we first describe the equations of the load shifting model (Section~\ref{sec_met_model}). Next, the determination of the input data for the model is presented, starting with the regionalisation of the energy demand and maximum capacity for all considered technologies (Section~\ref{sec_met_reg_endem_max_cap}). In Section~\ref{sec_met_sched_load} the calculation of the scheduled load is introduced. Subsequently, the determination of the remaining input parameters is explained: the time frame of management (Section~\ref{sec_met_timeframe}), the load increase and decrease shares (Section~\ref{sec_met_s_inc_dec}), the flexible share (Section~\ref{sec_flexible_share}) and the load shifting costs (Section~\ref{sec_costs}). While we refer to the 2018 status quo scenario in the previous sections, in Section~\ref{sec_met_fut_cen} the parameters for modelling the 2030 future scenario are presented.

\begin{table}[htbp]   
\begin{framed}
\small

\nomenclature[A]{AC}{Air conditioning}
\nomenclature[A]{BDEW}{German Association of Energy and Water Industries}
\nomenclature[A]{COP}{Coefficient of performance}
\nomenclature[A]{CTS}{Commercial, trade and services}
\nomenclature[A]{DHW}{Domestic hot water}
\nomenclature[A]{DR}{Demand response}
\nomenclature[A]{GVA}{gross value added}
\nomenclature[A]{VRE}{Variable renewable energy sources}
\nomenclature[A]{NUTS}{Nomenclature des Unités territoriales statistiques}
\nomenclature[A]{NUTS-0}{Country level}
\nomenclature[A]{NUTS-3}{Administrative district level (401 districts in Germany)}
\nomenclature[A]{PtG}{Power-to-Gas}
\nomenclature[A]{PtH}{Power-to-Heat}

\nomenclature[V,01]{$c$}{Load shifting technology index\nomunit{$\textendash$}}
\nomenclature[V,02]{$i$}{Administrative district index\nomunit{$\textendash$}}

\nomenclature[V,10]{$\Delta t$}{Time frame of management \nomunit{$h$}}
\nomenclature[V,11]{$L$}{Scheduled load\nomunit{$MW$}}
\nomenclature[V,12]{$\Lambda$}{Maximum capacity\nomunit{$MW$}}
\nomenclature[V,13]{$s_{dec/inc}$}{Load decrease / increase limit\nomunit{$\textendash$}}
\nomenclature[V,14]{$P_{min/max}$}{Maximum load decrease / increase\nomunit{$MW$}}
\nomenclature[V,15]{$E_{min/max}$}{Maximum energy postponing / preponing\nomunit{$MWh$}}

\nomenclature[V,20]{$s_{flex}$}{Flexible share\nomunit{$\textendash$}}
\nomenclature[V,21]{$s_{util}$}{Utilisation rate\nomunit{$\textendash$}}

\nomenclature[V,30]{$c_{inc}$}{Specific investment costs\nomunit{$\text{\euro{}}/MW$}}
\nomenclature[V,31]{$c_{fix}$}{Specific annual fixed costs\nomunit{$\text{\euro{}}/MW/a$}}
\nomenclature[V,32]{$c_{var}$}{Specific variable costs\nomunit{$\text{\euro{}}/MWh$}}

\nomenclature[V,40]{$x_{2030}$}{Relative change of installed capacity until 2030\nomunit{$\textendash$}}
\nomenclature[V,41]{$P_{2030}$}{Installed capacity in 2030\nomunit{$GW$}}


\printnomenclature
\end{framed}
\end{table}

\subsection{Load shifting model}\label{sec_met_model}
\begin{figure}[h!]
\centering\includegraphics[width=0.8\linewidth]{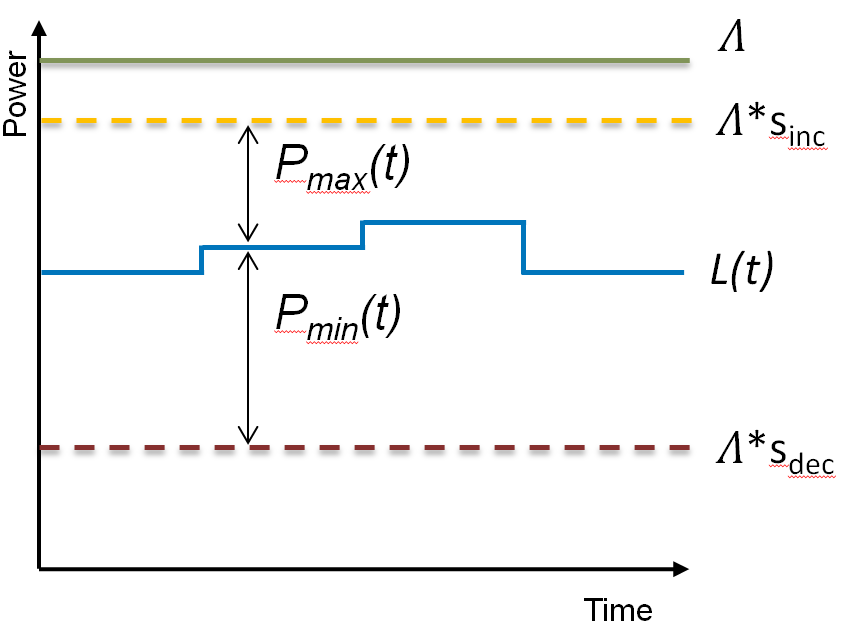}
\caption{Schematic illustration of the load increase and decrease buffers, $P_{max}$ and $P_{min}$.}
\label{fig_pmaxmin_schema}
\end{figure}

\begin{figure}[h!]
\centering\includegraphics[width=0.8\linewidth]{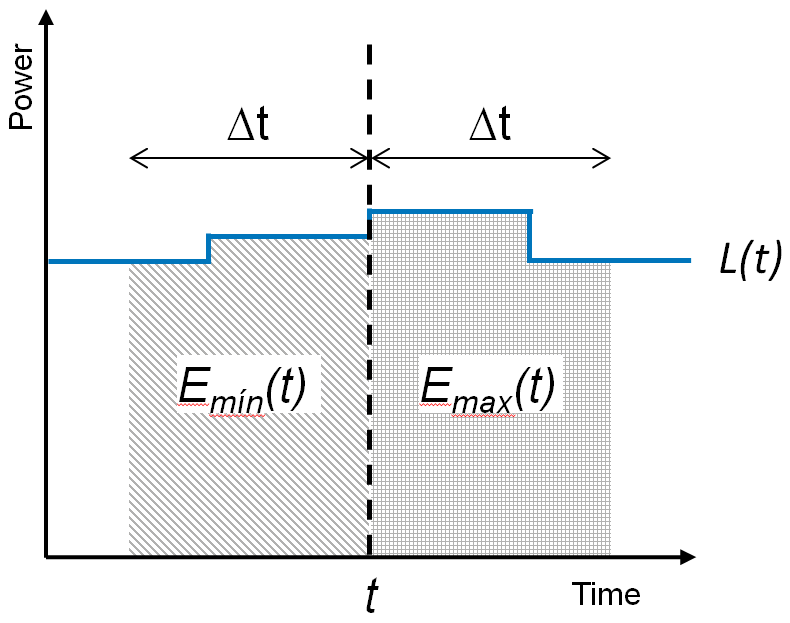}
\caption{Schematic illustration of the energy preponing and postponing buffers, $E_{max}$ and $E_{min}$.}
\label{fig_emaxmin_schema}
\end{figure}

In order to ease the integration into energy system models, in this work load shifting is modelled as an energy storage-equivalent operation. The model is an enhancement\footnote{The load increase and decrease parameters ($s_{inc}/s_{dec}$) were added in this work.} of a formulation that was developed at the DLR Institute of Networked Energy Systems in~\cite{kleinhans2014towards}. The following input parameters are used by the model for each regarded category, $c$, of load shifting technologies:
\begin{enumerate}
    \item Scheduled load $L^c(t)$: Load time series for a given application, without any load shifting modifications.
    
    \item Time frame of management $\Delta t^c$: Maximum duration by which loads can be postponed or preponed.
    
    \item Maximum capacity $\Lambda^c$: Installed capacity of a given application.
    
    \item Load increase and decrease limits, $s_{inc}^c$ and $s^c_{dec}$: Share of the maximum capacity up to which the load can be increased or decreased.
\end{enumerate}
The load shifting process of the scheduled load $L^c(t)$ results in a new time series, the realised load $R^c(t)$. In terms of load shifting as an energy storage-equivalent operation, the charging rate of the storage can be defined as:
\begin{equation}
P^c(t)=R^c(t)-L^c(t),
\end{equation}
where the storage is charged for $P(t)>0$ and discharged for $P(t)<0$. Integrating the charging rate over time yields the filling level $E(t)$ of the storage, 
\begin{equation}
E^c(t)=\int_{0}^{t} P^c(t')~dt'.
\end{equation}
Using this terminology storage-equivalent buffers can be defined as: 
\begin{align}
&E_{max}^c(t)=\int_{t}^{t+\Delta t} L^c(t')~dt',\\ \label{eq_e_max}
&E_{min}^c(t)=-\int_{t-\Delta t}^{t} L^c(t')~dt',\\ \label{eq_e_min}
&P_{max}^c(t)= \Lambda^c \cdot s_{inc}^c - L^c(t),\\
&P_{min}^c(t)= - (L^c(t) - \Lambda^c \cdot s_{dec}^c).
\end{align}
These buffers serve as boundary conditions for the charging rate and filling level of the storage:
\begin{align}
&E_{max}^c(t)\leq E^c(t) \leq E_{min}^c(t)~\forall~t,\\
&P_{max}^c(t)\leq P^c(t) \leq P_{min}^c(t)~\forall~t.
\end{align}
The load increase and decrease buffers, $P_{max}$ and $P_{min}$, and the energy preponing and postponing buffers, $E_{max}$ and $E_{min}$, are schematically illustrated in Figure~\ref{fig_pmaxmin_schema} and~\ref{fig_emaxmin_schema}. As these buffers can be determined before the load shifting dispatch optimisation, the presented model allows for a computationally efficient integration of load shifting into energy system models.

The cumulated time series of all technologies for the scheduled and realised load, as well as the buffers, are obtained by summing over all categories $c$.  
In the following sections we describe the determination of the input data for the load shifting model and the utilised input parameters are summarised in Table~\ref{tab_dsm_parameters}.

\subsection{Regionalisation of the energy demand and maximum capacity}\label{sec_met_reg_endem_max_cap}


\begin{table*}[h!]
\caption{Input parameters for the energy demand regionalisation for 2018. 
Additional input parameters are given in previous works of the authors on the regionalisation of the residential heat demand~\cite{heitkoetter2020regionalised} and industrial energy demand~\cite{schmidt2019nuts}.}
\label{tab_param_endem_reg}
\scriptsize
\begin{center}
\begin{tabularx}{\textwidth}{l l l l l l}
\hline
Sector&Symbol&Value&Unit&Description&Source\\
\hline
Residential&$\bar{E}^{res}$&129&TWh& Annual electricity demand of the German residential sector&\cite{bmwi2018energiedaten}\\
&$x^{res,wd}$&0.09&\textendash&Share of washing and drying machines in the residential electricity demand&\cite{klobasa2007dynamisch}\\
&$x^{res,ff}$&0.17&\textendash&Share of fridges and freezers in the residential electricity demand&\cite{klobasa2007dynamisch}\\
CTS&$\bar{E}^{cts}$&147&TWh&Annual electricity demand of the German CTS sector&\cite{bmwi2018energiedaten}\\
&$x^{cts,ve}$&0.12&\textendash&Share of ventilation in the CTS electricity demand&\cite{klobasa2007dynamisch}\\
&$x^{cts,ac}$&0.08&\textendash&Share of air conditioning in the CTS electricity demand&\cite{klobasa2007dynamisch}\\
&$x^{cts,co}$&0.02&\textendash&Share of cooling in the CTS electricity demand&\cite{klobasa2007dynamisch}\\
Industry&$\bar{E}^{ind}$&239&TWh&Annual electricity demand of the German industry sector&\cite{destatis2018stromverbrauch}\\
&$x^{ind,ve}$&0.017&\textendash&Share of ventilation in the CTS electricity demand&\cite{klobasa2007dynamisch}\\
&$x^{ind,co}$&0.022&\textendash&Share of cooling in the industrial electricity demand&\cite{icf2015study}\\
Power-to-heat&${<}COP^{hp}{>}$&2.89&\textendash&Annual average coefficient of performance of heat pumps&\cite{pellinger2016merit}\\
&${<}COP^{rs}{>}$&1&\textendash&Annual average coefficient of performance of resistive heaters&\cite{heitkoetter2020regionalised}\\
&$\bar{Q}^{cts}$&216&TWh&Annual heat demand of the German CTS sector&\cite{bmwi2018energiedaten}\\
&$x^{cts,hp}$&0.006&\textendash&Share of the CTS heat demand covered by heat pumps&\cite{ageb2017satellitenbilanz}\\
&$x^{cts,rs}$&0.10&\textendash&Share of the CTS heat demand covered by resistive heaters&\cite{bmwi2018energiedaten}\\
E-mobility&$e^{ev}$&2.8&MWh&Annual electricity demand per electric vehicle&\cite{nitsch2012langfristszenarien}\\
&$n^{ev}$&150000&\textendash&Number of plug-in electric vehicles in Germany&\cite{kba2019bestand}\\
\hline

\hline

\end{tabularx}%
\label{tab_input_param_values}
\end{center}
\end{table*}

The maximum capacity $\Lambda_i^c$ per NUTS-3 district $i$ of each load shifting technology is determined by using the NUTS-3 annual energy demand $E_i^c$ and the average annual utilisation rate $s_{util}^c$:
\begin{equation}
\Lambda_i^c = \frac{E_i^c \cdot s_{flex}^c}{8760~h  \cdot s_{util}^c} \label{eq_lamda}
\end{equation}
The flexible share parameter, $s_{flex}^c$, accounts for socio-technical restrictions (see Section~\ref{sec_flexible_share}). 
In the next subsections the NUTS-3 annual energy demand is derived for the different demand sectors and technologies. Table~\ref{tab_param_endem_reg} gives an overview of the numeric values of the utilised parameters for the energy demand regionalisation.

\subsubsection{Residential sector}
In the residential sector washing and drying machines, as well as fridges and freezers are suitable for load shifting~\cite{klobasa2007dynamisch}. Therefore, we multiplied the annual electricity demand of the German residential sector, $\bar{E}^{res}$, with the share of washing and drying machines, $x^{res,wd}$, and fridges and freezers, $x^{res,ff}$. Hence, the NUTS-3 annual energy demand, $E_i^{res, wd/ff}$, was determined via:
\begin{equation}
    E_i^{res, wd/ff} = x_i^{res} \cdot x^{res,wd/ff} \cdot \bar{E}^{res},
\end{equation}
where ${x}_i^{res}$ is the share of each administrative district in the German residential electricity demand. The electricity demand per resident rises with higher income and fewer members per household~\cite{pellinger2016merit}. Thus ${x}_i^{res}$ was computed by:
\begin{equation}
    x_i^{res} = (I_i \sum_{sc=1}^{sc} \omega_{sc} N_{i,sc}) / \bar{I}.
\end{equation}
Therein $I_i$ is the average income per inhabitant per district and $\bar{I}$ the cumulated income of all residents of Germany~\cite{destatis2017einkommen}.
In order to account for the increasing electricity demand per household member of households with fewer members~\cite{oberascher2016verbrauch}, we considered six household size classes $m_{sc} \in \{1, 2, 3, 4, 5, 6 \}$. Next, the weighting factor $\omega_{sc}$ was defined by,
\begin{equation}
    \omega_{sc} = 0.75 \cdot m_{sc} + 1.5,
\end{equation}
and multiplied with the number of households per household class in each administrative district, $N_{i,sc}$. 

\subsubsection{Commercial, trade and services sector}
In the commercial, trade and services (CTS) sector cross-sectional technologies that are used in multiple CTS branches provide the highest load shifting potential~\cite{klobasa2007dynamisch}.
Out of the cross-sectional technologies, ventilation and air conditioning appliances, as well as cooling processes are considered as suitable for load shifting~\cite{klobasa2007dynamisch}. We multiplied the overall German CTS electricity demand, $\bar{E}^{cts}$, with the share that is needed for ventilation, $x^{cts,ve}$, for air conditioning, $x^{cts,ac}$, and for cooling, $x^{cts,co}$~\cite{icf2015study}. To determine the NUTS-3 distribution of the regarded processes energy demand, $E_i^{cts,ve/ac/co}$, the share of the CTS employment of the respective administrative district, $x_i^{cts}$~\cite{destatis2018erwerbstaetige}, in the overall German CTS employment was used:   
\begin{equation}
E_i^{cts, ve/ac/co} = \bar{E}^{cts} \cdot  x^{cts,ve/ac/co} \cdot x_i^{cts} \label{eq_e_cts_cross}
\end{equation}

\subsubsection{Industrial sector}
As like as in the CTS sector, also in the industrial sector, non-process relevant ventilation appliances and cooling processes in the food industry are considered as suitable for load shifting~\cite{klobasa2007dynamisch}. The methodology for determining these load shifting potentials is equivalent to Eq.~\ref{eq_e_cts_cross}, but instead uses the respective energy demand and demand shares in the industry,$\bar{E}^{ind}$, $x^{ind,ve}$, $x^{ind,co}$ and $x_i^{ind}$~\cite{destatis2018betriebe,destatis2019stromverbrauch}.\footnote{For more information refer to the supplementary material.}

Furthermore, the following energy intensive industrial processes are suitable for load shifting~\cite{gils2015balancing}: cement milling, mechanical wood pulping, paper production, recycled paper pulping, and air separation.
In~\cite{schmidt2019nuts} the authors of this study show that it leads to significant errors, when the energy intensive processes are regionalised using statistical NUTS-3 employment or GVA data. Instead, we identified individual industrial plants, as well as their locations, and production capacities, $c_{pl}$, by using registers of German national industry associations. The production capacities were multiplied with the specific energy demand $e_{pl}$ and the utilisation factor, $s_{util,pl}$. All plants per administrative district $i$ were summed to determine the annual energy demand of the energy intensive processes,
\begin{equation}
E_{i}^{ind, int} = \sum_{pl} c_{pl} \cdot e_{pl} \cdot s_{util,pl}. \label{eq_ind_plant_sp}
\end{equation}
The plant specific regionalisation methodology is described in detail in~\cite{schmidt2019nuts} and the results are published as an open dataset.\footnote{The open dataset is available at: \href{https://doi.org/10.5281/zenodo.3613766}{https://doi.org/10.5281/zenodo.3613766}}

\subsubsection{Power-to-heat}\label{reg_cap_pth}
In order to assess the role of sector coupling technologies for load shifting, we clustered the electric heating technologies from the residential, CTS and industry sector as one separate power-to-heat sector.

\subsubsection*{Residential power-to-heat}\label{sec_res_pth}
The regionalisation of the residential power-to-heat load is based on an open dataset developed in a previous work of the authors~\cite{heitkoetter2020regionalised}.\footnote{The open dataset is available at:~\href{https://doi.org/10.5281/zenodo.2650200}{https://doi.org/10.5281/zenodo.2650200}}
Using a special evaluation of census enumeration data, 729 residential building categories $b$ were defined. An area specific annual heat demand $q_b''$ was assigned to each building category, depending on the year of construction of the buildings, the type of building, number of flats per building, floor area, heating type and number of residents per building. To yield the absolute annual heat demand, the area specific heat demand was multiplied with the average floor area $A_b$ of the buildings in the respective category, as well as the number of buildings per category per administrative district $n_b$. To calculate the electrically-covered heat demand, the share of buildings was taken into account that are equipped with heat pumps, $x_b^{hp}$, and resistive heating devices $x_b^{rs}$. The demand of all building categories b was summed, yielding the overall demand per administrative district. 
The annual electricity demand $E_i^{hp/rs}$ was determined by dividing the heat demand by the average annual coefficient of performance ${<}COP^{hp/rs}{>}$~\cite{heitkoetter2020regionalised}:
\begin{equation}
E_i^{sh,hp/rs} = {<}COP^{hp/rs}{>}^{-1} \cdot \sum_b q_b'' \cdot A_b \cdot n_b \cdot x_b^{hp/rs} \label{eq_res_an_heat_dem}
\end{equation}

Next, the load shifting potential of electric domestic hot water (DHW) heaters was determined. An average annual demand per person $q^{dhw}$ was assigned and multiplied with the number of residents $n_b^{res}$ in the respective building category. As defined in Eq.~\ref{eq_res_an_heat_dem} it was summed over all building categories and the heat demand was converted to electric energy demand:
\begin{equation}
E_i^{dhw,hp/rs} = {<}COP^{hp/rs}{>}^{-1}  \sum_b q^{dhw} \cdot n_b^{res} \cdot n_b \cdot x_b^{hp/rs}. \label{eq_res_an_dhw_dem}
\end{equation}

The installed capacities of centralised PtH facilities in district heating grids are given in~\cite{christidis_2017_eneff}. We assigned these capacities to the respective administrative districts, in which the district heating grids are located. According to~\cite{christidis_2017_eneff}, only resistive heaters are installed in German district heating grids, while there are no large-scale heat pumps.  

\subsubsection*{CTS power-to-heat}
For the regional distribution of commercial building types, comprehensive statistical data such as for residential buildings does not exist. We therefore regionalised the overall German annual heat demand in the CTS sector, $\bar{Q}^{cts}$, to the NUTS-3 level, according to the share of the NUTS-3 annual gross value added (GVA), $x_{i}^{gva}$~\cite{destatis2017bruttoinlandsprodukt}, in the total German GVA. The CTS heat demand was converted to electricity demand, $E_i^{cts,hp/rs}$, using the demand share covered by heat pumps $x^{cts,hp}$~\cite{ageb2017satellitenbilanz} and by resistive heating $x^{cts,rs}$~\cite{bmwi2018energiedaten}, as well as the respective annual average coefficients of performance ${<}COP^{hp/rs}{>}$:
\begin{equation}
E_i^{cts,hp/rs} = \bar{Q}^{cts} \cdot x_{i}^{gva} \cdot x^{cts,hp/rs} * {<}COP^{hp/rs}{>}^{-1}. \label{eq_cts_an_dhw_dem}
\end{equation}

\subsubsection*{Industrial power-to-heat}
The regionalisation of the industrial process heat demand is based on a previous work of the authors~\cite{schmidt2019nuts} and is summarised in this section. 
Amongst others, there is a significant process heat demand in the following industries: metal, minerals, mining, food, textile, paper, machinery and wood~\cite{schmidt2019nuts}.  
The annual NUTS-0 primary energy demand of the respective processes $\bar{E}^{ind,pri}_{br}$~\cite{schmidt2019nuts} was multiplied with the shares of the process heat $x_{br}^{ph}$~\cite{schmidt2019nuts}. Next, it was multiplied by the average conversion efficiency to useful heat energy $\eta_{br}^{ind,ph}$~\cite{schmidt2019nuts} per industrial branch, which depends on the used primary energy carriers, e.g. coal, oil or natural gas.
The NUTS-0 process heat demand was distributed to the administrative districts using the share of the NUTS-3 employment in the considered branches $x_{i,br}$~\cite{schmidt2019nuts} in the total German employment in those branches. We summed over all branches and divided the result by the average coefficient of performance, ${<}COP^{rs}{>}$, to yield the electric energy demand for covering industrial process heat in each administrative district,
\begin{equation}
E_{i} = {<}COP^{rs}{>}^{-1} \sum_{br} \bar{E}^{ind,pri}_{br} \cdot x_{br}^{ph} \cdot \eta_{br}^{ind,ph} \cdot x_{i,br}. \label{eq_en_dem_ph}
\end{equation}
There were no statistical data on the number of installed heat pumps in industrial facilities available. Therefore, we assumed that all industrial PtH is covered by resistive heaters. 

\subsubsection{E-mobility}
Next, we determined the annual energy demand of electric vehicles, $E_{i}^{ev}$, in all German administrative districts. To calculate this, the annual electrical energy demand per electric vehicle, $e^{ev}$, was multiplied with the number of plug-in electric vehicles in Germany $n^{ev}$ and the share of registered electric vehicles per administrative district, $x_i^{ev}$~\cite{kba2019bestand}, in all electric vehicles in Germany:
\begin{equation}
E_{i}^{ev} =  e^{ev} \cdot n^{ev} \cdot x_i^{ev}. \label{eq_en_dem_ev}
\end{equation}
In contrast to the other considered technologies, the maximum capacity is not constant for e-mobility, as the share of the connected electric vehicles, $x_t^{ev,conn}$, varies during the day. Thus the temporally resolved maximum capacity $\Lambda_t^{ev}$ is calculated by,
\begin{equation}
\Lambda_t^{ev} =  \Lambda^{ev} \cdot x_t^{ev,conn}. \label{eq_ev_lamda_t}
\end{equation}
The values of $x_t^{ev}$ are provided in~\cite{luca2014large} and were derived from the German mobility statistics~\cite{infas2003mobilitaet}.

\subsubsection{Power-to-gas}
For the NUTS-3 regionalisation of power-to-gas capacities $\Lambda_i^{hy/me}$, we assigned the locations of the PtG plants listed in~\cite{thema2019power} to the respective administrative districts they are located in. We differentiated between capacities of power-to-hydrogen plants, $c_{pl}^{hy}$ and power-to-methane plants, $c_{pl}^{me}$ and summed the capacities per district:

\begin{equation}
\Lambda_i^{hy/me} = \sum_{pl} \Lambda_{pl}^{hy/me}. \label{eq_cap_ptg}
\end{equation}

\subsection{Scheduled load}\label{sec_met_sched_load}
The scheduled load time series, $L_{i}(t)$, for each administrative district and each technology were calculated via the annual energy demand, the flexible share, the share of the energy demand in each time step, $x_L(t)$, and the time step length, $\delta t$:
\begin{equation}
L_i(t) = \frac{E_i \cdot s_{flex} \cdot x_L(t)}{\delta t}.\label{eq_l_i_t}
\end{equation}
The values for $x_L(t)$ were determined using normalised daily load profiles, as well as additional yearly load profiles for the PtH and cement milling technologies. In the following sections, the applied methodology is summarised and the resulting load profiles are visualised in Figure~\ref{temporal_distribution}. For further information and the numerical values of the load profiles, refer to the given sources and the supplementary material of this manuscript.

\subsubsection{Daily load profiles}

A commonly used reference for daily load profiles are the standard load profiles of the German Association of Energy and Water Industries (BDEW)~\cite{funfgeld2000anwendung}. We used the BDEW standard load profile for households, $H0$, for modelling the load profile of the residential appliances. The considered cross-sectional technologies in the CTS sector, cooling, ventilation and AC, have high utilisation rates. Therefore the BDEW standard load profile $G3$ for commercial consumers with high full load hours was applied.

There were no BDEW standard load profiles found that were broken down on the considered industrial processes. Following~\cite{gils2015balancing}, we assumed constant daily load profiles for the energy intensive industry processes. Industrial ventilation processes were also assumed to have a constant load profile, while a weekend decline of $40\%$ on Saturdays and $50\%$ on Sundays was taken into account~\cite{gils2015balancing}. The industrial cooling load was estimated to be $50\%$ reduced during morning peak load hours, $5\%$ reduced on Saturdays and $10\%$ reduced on Sundays~\cite{gils2015balancing}.

Concerning residential PtH, we used daily load profiles that were derived by the authors of this paper in~\cite{heitkoetter2020regionalised}, based on long term measurements~\cite{lange_2018_novaref}. The data were obtained measuring the heat flow and return at the heat exchangers in residential buildings and thus represent the actual demand of the residents. For PtH in the CTS sector, we used the BDEW heat demand standard daily load profile, $GHD$, for commercial, trade and services~\cite{bdew2011abwicklung}. For PtH covering industrial process heat the load profiles were adopted from~\cite{gils2015balancing} and take into account demand shares and typical full load hours of different manufacturing branches.

Charging of electric vehicles was assumed to follow the load profile developed in~\cite{luca2014large}. The profile was derived from German mobility statistics data~\cite{infas2003mobilitaet} and contains a morning and evening charging peak. Power-to-gas plants were treated as energy intensive industrial processes and the load profile was approximated to be constant.

\subsubsection{Yearly load profiles}
Out of the considered technologies only PtH for space heating and cement production are assumed to have a varying load profile during the course of the year, because these processes are dependent on the ambient temperature. 
For modelling the yearly load profile of PtH in the residential sector, the methodology developed by the authors of this paper in~\cite{heitkoetter2020regionalised} was adopted. The space heating load $\dot{Q}_{b,t}$ at each point of time $t$ depends on the difference between the ambient temperature $T_t^{amb}$ and the heating limit temperature $T^{hl}$, as well as on the ambient temperature specific heat demand factor $H$. Both, $T_b^{hl}$ and $H_b$ depend on the building attributes, e.g. insulation quality or floor area. Therefore individual $T_b^{hl}$ and $H_b$ values were assigned for each residential building category~\cite{heitkoetter2020regionalised}, which were introduced in Section~\ref{sec_res_pth}. The heat load for space heating, $\dot{Q}_{b,t}$, could thus be calculated, as follows:  
\begin{align}
T^{amb} \geq T^{hl}: \dot{Q}_{b,t} &= 0 \label{eq_q_dot_ta_gr_thl},\\
T^{amb} < T^{hl}: \dot{Q}_{b,t} &= H_b \cdot (T_b^{hl} - T_t^{amb})  \label{eq_q_dot_ta_sm_thl}.
\end{align}

For $T_t^{amb}$ we used the daily average ambient temperatures from 2018, measured at the closest weather station to the respective administrative district, which were obtained from the website of the German Meteorological Service~\cite{dwd2018cdc}. 
Due to the thermal inertia of the building mass, the temperatures of previous days influence the daily heat load~\cite{hellwig2003entwicklung}. To account for this influence, we applied a geometric series to the temperature input data, considering the three previous days, as described in detail in~\cite{heitkoetter2020regionalised}.

Further, $\dot{Q}_{b,t}$ was multiplied with the number of buildings per building category and administrative district, as well the share of buildings equipped with heat pumps and resistive space heating. The result was summed over all building categories and divided by the temporally resolved coefficient of performance, yielding the scheduled electric load for space heating per administrative district: 
\begin{equation}
P_{i,t}^{sh,hp/rs}= (COP_t^{hp/rs})^{-1}  \sum_b \dot{Q}_{b,t} \cdot n_b \cdot x_b^{hp/rs}
\end{equation}
While the coefficient of performance for resistive space heating, $COP_t^{rs}$, was assumed to be constant, the coefficient of performance of heat pumps, $COP_t^{hp}$, depends on the ambient temperature. Therefore, we adopted an approach by Ruhnau~\cite{ruhnau2019time}, who calculates the COP time series of air sourced (ASHP), ground sourced (GSHP) and water sourced (WSHP) heat pumps\footnote{For more information on heat pump types, refer to~\cite{staffell2012review}.} based on a quadratic regression of manufacturer data:
\begin{equation}
COP_t^{hp}=\left\{
\begin{matrix}
6.08-0.09 \cdot \Delta T + 0.0005\cdot \Delta T^2,~ASHP, \\
10.29-0.21 \cdot \Delta T + 0.0012\cdot \Delta T^2,~GSHP,\\
9.97-0.20 \cdot \Delta T + 0.0012\cdot \Delta T^2,~WSHP.
\end{matrix}
\right.
\end{equation}
Therein $\Delta T$ comprises all possible combinations of source and sink temperatures, 
\begin{align}
\Delta T^{sink,source} &= T^{sink} - T^{source},
\end{align}
where $source~\in~\{air, ground, water\}$ and $sink~\in~\{radiator~heating, floor~heating\}$. For $T^{source}$, hourly measured air and ground temperatures, measured at the closest weather station to the respective administrative district, were obtained from the German Meteorological Service~\cite{dwd2018cdc}. For WSHP a constant ground water temperature of $10^{\circ}C$ throughout the year was assumed, following~\cite{ruhnau2019time}.
The sink temperatures depend on the utilised sink type, as well as the ambient temperature. For the calculation of $T^{sink}$, as well as the assumed heat transfer temperature differences, refer to the supplementary material of this paper or the description of the methodology given in~\cite{ruhnau2019time} that was adopted here.
Subsequently, a weighted average of COP time series for the different technologies was calculated using the shares of installed heat pumps in Germany, $55\%$ ASHP, $39\%$ GSHP and $6\%$ WSHP~\cite{bwp_waermepump_2016, bwp2018branchenstudie}.

As introduced in Section~\ref{reg_cap_pth}, no data were available for a detailed regionalisation of commercial building types. We therefore used the BDEW annual heat demand load profile~\cite{bdew2011abwicklung} for modelling the influence of the ambient temperature, which is an aggregated load profile for all CTS building types. The same regionalised ambient temperature time series were used as for the residential buildings. For converting thermal demand to electric demand we used the same methodology as described above for residential PtH.

The load profile of cement mills varies during the course of the year, because low temperatures in winter prevent construction work. Following~\cite{schmidt2019nuts}, we used $1.5^\circ C$ as threshold temperature, below which cement mills are assumed to be shut down and their load profile is set to zero.

\subsection{Time frame of management}\label{sec_met_timeframe}
The time frame of management parameter, $\Delta t^c$, specifies the maximum duration by which loads can be preponed or postponed and mainly depends on the storage capacity of the considered processes~\cite{gils2015balancing}. For heating applications the thermal inertia of the building and the size of the hot water storage are the determining factors, for industrial processes the product storage capacity is decisive and for electric vehicles the installed battery capacity. Residential washing and drying appliances are the only considered processes that do not have a physical storage capacity. The time frame of management parameter specifies for how long these processes can be shifted without significantly disturbing the user. In Table~\ref{tab_dsm_parameters}, the numerical values for $\Delta t$ are provided, as well as the respective literature sources.

\begin{table*}[htbp]
\centering
 \caption{Input parameters for modelling the load shifting process, the future scenario, investment costs, fixed costs and variable costs; \textsuperscript{$*$}own assumption; \textsuperscript{$\dagger$}averaged from~\cite{gils2015balancing} and \cite{steurer2017analyse}; \textsuperscript{$\P$}approx. average from industrial processes; \textsuperscript{$\P\P$}adopted from residential heat pumps; \textsuperscript{$\S$}adopted from decentralised residential heating; \textsuperscript{$\S\S$}adopted from power-to-methane;
 \textsuperscript{$\ddag$}adopted from air separation.}
 \label{tab_dsm_parameters}
 \scriptsize
  \begin{tabular}{l l l l l l l l l l l}
  \hline
  Sector & Technology, c & $\Delta t^c~[h]$ & $s_{util}^c~[\textendash]$& $s_{dec}^c~[\textendash]$ & $s_{inc}^c~[\textendash]$ & $s_{flex}^c~[\textendash]$ & $x_{2030}^c/P_{2030}^c$ & $c_{inv}^c[\text{\euro{}}/MW]$ & $c_{fix}^c[\text{\euro{}}/MW/a]$ & $c_{var}^c[\text{\euro{}}/MWh]$ \\
  \hline
  Residential & Washing,drying & 6\textsuperscript{\cite{gils2015balancing}} & 0.01\textsuperscript{\cite{steurer2017analyse}} & 0.0025$\dagger$ & 0.025$\dagger$ & 0.4\textsuperscript{\cite{steurer2017analyse}} & 0.65 [-]\textsuperscript{\cite{steurer2017analyse}} & 220000\textsuperscript{$*$;\cite{pellinger2016merit}} & 42000\textsuperscript{$*$;\cite{pellinger2016merit}} & 50\textsuperscript{\cite{gils2015balancing}} \\
   & Cooling,freezing & 2\textsuperscript{\cite{gils2015balancing}} & 0.33\textsuperscript{\cite{klobasa2007dynamisch}} & 0\textsuperscript{\cite{steurer2017analyse}} & 1\textsuperscript{\cite{steurer2017analyse}} & 0.4\textsuperscript{\cite{steurer2017analyse}} & 0.7 [-]\textsuperscript{\cite{steurer2017analyse}} & 220000\textsuperscript{$*$;\cite{pellinger2016merit}} & 42000\textsuperscript{$*$;\cite{pellinger2016merit}} & 50\textsuperscript{$*$;\cite{gils2015balancing}} \\
  CTS & Cooling,ventilation,AC & 1\textsuperscript{\cite{gils2015balancing}} & 0.67\textsuperscript{\cite{gils2015balancing}} & 0\textsuperscript{\cite{steurer2017analyse}} & 1\textsuperscript{\cite{steurer2017analyse}} & 0.5\textsuperscript{\cite{steurer2017analyse}} & 1.5 [-]\textsuperscript{$*$;\cite{steurer2017analyse}} & 10000\textsuperscript{\cite{gils2015balancing}} & 300\textsuperscript{\cite{gils2015balancing}} & 5\textsuperscript{\cite{gils2015balancing}} \\
  Industry & Air separation & 4\textsuperscript{\cite{klobasa2007dynamisch}} & 0.86\textsuperscript{\cite{steurer2017analyse}} & 0.4\textsuperscript{\cite{steurer2017analyse}} & 0.95\textsuperscript{\cite{steurer2017analyse}} & 0.3\textsuperscript{\cite{klobasa2007dynamisch}} & 0.9 [-]\textsuperscript{\cite{steurer2017analyse}} & 200\textsuperscript{\cite{steurer2017analyse}} & 100\textsuperscript{\cite{steurer2017analyse}} & 150\textsuperscript{\cite{steurer2017analyse}} \\
   & Cement & 4\textsuperscript{\cite{klobasa2007dynamisch}} & 0.65\textsuperscript{\cite{steurer2017analyse}} & 0\textsuperscript{\cite{steurer2017analyse}} & 0.95\textsuperscript{$\P$} & 0.61\textsuperscript{\cite{steurer2017analyse}} & 0.93 [-]\textsuperscript{\cite{steurer2017analyse}} & 1500\textsuperscript{\cite{steurer2017analyse}} & 19100\textsuperscript{\cite{steurer2017analyse}} & 200\textsuperscript{\cite{steurer2017analyse}} \\
   & Pulp & 2\textsuperscript{\cite{klobasa2007dynamisch}} & 0.83\textsuperscript{\cite{steurer2017analyse}} & 0\textsuperscript{\cite{steurer2017analyse}} & 0.95\textsuperscript{\cite{steurer2017analyse}} & 0.7\textsuperscript{\cite{steurer2017analyse}} & 0.96 [-]\textsuperscript{\cite{steurer2017analyse}} & 2300\textsuperscript{\cite{steurer2017analyse}} & 2000\textsuperscript{\cite{steurer2017analyse}} & 250\textsuperscript{\cite{steurer2017analyse}} \\
   & Paper & 3\textsuperscript{\cite{gils2015balancing}} & 0.86\textsuperscript{\cite{steurer2017analyse}} & 0\textsuperscript{\cite{steurer2017analyse}} & 0.95\textsuperscript{\cite{steurer2017analyse}} & 0.15\textsuperscript{\cite{steurer2017analyse}} & 1.17 [-]\textsuperscript{\cite{steurer2017analyse}} & 2300\textsuperscript{\cite{steurer2017analyse}} & 2000\textsuperscript{\cite{steurer2017analyse}} & 200\textsuperscript{\cite{steurer2017analyse}} \\
   & Recycled paper & 3\textsuperscript{\cite{gils2015balancing}} & 0.85\textsuperscript{\cite{steurer2017analyse}} & 0\textsuperscript{\cite{steurer2017analyse}} & 0.95\textsuperscript{\cite{steurer2017analyse}} & 0.7\textsuperscript{\cite{steurer2017analyse}} & 1.25 [-]\textsuperscript{\cite{steurer2017analyse}} & 2300\textsuperscript{\cite{steurer2017analyse}} & 2000\textsuperscript{\cite{steurer2017analyse}} & 100\textsuperscript{\cite{steurer2017analyse}} \\
   & Cooling & 2\textsuperscript{\cite{gils2015balancing}} & 0.67\textsuperscript{\cite{gils2015balancing}} & 0.5\textsuperscript{\cite{gils2015balancing}} & 0.9\textsuperscript{\cite{gils2015balancing}} & 0.63\textsuperscript{\cite{steurer2017analyse}} & 1 [-]\textsuperscript{\cite{gils2015balancing}} & 5000\textsuperscript{\cite{gils2015balancing}} & 150\textsuperscript{\cite{gils2015balancing}} & 20\textsuperscript{\cite{gils2015balancing}} \\
   & Ventilation & 1\textsuperscript{\cite{gils2015balancing}} & 0.8\textsuperscript{\cite{gils2015balancing}} & 0.5\textsuperscript{\cite{gils2015balancing}} & 1\textsuperscript{\cite{gils2015balancing}} & 0.5\textsuperscript{\cite{steurer2017analyse}} & 1 [-]\textsuperscript{\cite{gils2015balancing}} & 10000\textsuperscript{\cite{gils2015balancing}} & 300\textsuperscript{\cite{gils2015balancing}} & 5\textsuperscript{\cite{gils2015balancing}} \\
  PtH & Process heat (ind) & 3\textsuperscript{$\P$} & 0.8\textsuperscript{$\P$} & 0.2\textsuperscript{$\P$} & 0.95\textsuperscript{$\P$} & 0.5\textsuperscript{$\P$} & 2.9[-]\textsuperscript{$\P\P$} & 3500\textsuperscript{$\P$} & 3600\textsuperscript{$\P$} & 100\textsuperscript{$\P$} \\
   & Heat pumps (res) & 3\textsuperscript{\cite{steurer2017analyse}} & 0.22\textsuperscript{\cite{heitkoetter2020regionalised}} & 0\textsuperscript{\cite{steurer2017analyse}} & 0.75\textsuperscript{\cite{steurer2017analyse}} & 0.4\textsuperscript{\cite{steurer2017analyse}} & 2.9[-]\textsuperscript{\cite{nep2019szenariorahmen}} & 62000\textsuperscript{$*$;\cite{pellinger2016merit}} & 12000\textsuperscript{$*$;\cite{pellinger2016merit}} & 10\textsuperscript{$*$;\cite{gils2015balancing}} \\
   & Resistive sh. (res) & 12\textsuperscript{\cite{gils2015balancing}} & 0.22\textsuperscript{\cite{heitkoetter2020regionalised}} & 0\textsuperscript{\cite{steurer2017analyse}} & 0.75\textsuperscript{\cite{steurer2017analyse}} & 0.4\textsuperscript{\cite{steurer2017analyse}} & 1 [-]\textsuperscript{\cite{heitkoetter2020regionalised}} & 39200\textsuperscript{$*$;\cite{pellinger2016merit}} & 7000\textsuperscript{$*$;\cite{pellinger2016merit}} & 10\textsuperscript{$*$;\cite{gils2015balancing}} \\
   & Resistive dhw. (res) & 12\textsuperscript{\cite{gils2015balancing}} & 0.03\textsuperscript{\cite{gils2015balancing}} & 0\textsuperscript{\cite{steurer2017analyse}} & 0.17\textsuperscript{\cite{steurer2017analyse}} & 0.25\textsuperscript{\cite{apel2012notwendiger}} & 1 [-]\textsuperscript{\cite{heitkoetter2020regionalised}} & 155000\textsuperscript{$*$;\cite{pellinger2016merit}} & 29500\textsuperscript{$*$;\cite{pellinger2016merit}} & 10\textsuperscript{$*$;\cite{gils2015balancing}} \\
   & PtH in district heating & 12\textsuperscript{$\S$} & 0.22\textsuperscript{\cite{heitkoetter2020regionalised}} & 0\textsuperscript{$\S$} & 0.95\textsuperscript{$\P$} & 0.4\textsuperscript{$S$} & 1.9 GW\textsuperscript{\cite{heitkoetter2020regionalised}} & 200\textsuperscript{$\ddag$} & 100\textsuperscript{$\ddag$} & 10\textsuperscript{$\P$} \\
   & Heat pumps (cts) & 3\textsuperscript{\cite{steurer2017analyse}} & 0.22\textsuperscript{\cite{heitkoetter2020regionalised}} & 0\textsuperscript{\cite{steurer2017analyse}} & 0.75\textsuperscript{\cite{steurer2017analyse}} & 0.4\textsuperscript{\cite{steurer2017analyse}} & 2.9 [-]\textsuperscript{$\P\P$} & 20000\textsuperscript{$*$;\cite{gils2015balancing}} & 600\textsuperscript{$*$;\cite{gils2015balancing}} & 10\textsuperscript{$*$;\cite{gils2015balancing}} \\
  PtG & Power-to-methane & 24\textsuperscript{\cite{albrecht2016kommerzialisierung}} & 0.23\textsuperscript{\cite{nep2019szenariorahmen}} & 0\textsuperscript{$\P$} & 0.95\textsuperscript{$\P$} & 1\textsuperscript{$*$} & 0.4 GW\textsuperscript{\cite{nep2019szenariorahmen}} & 200\textsuperscript{$\ddag$} & 100\textsuperscript{$\ddag$} & 150\textsuperscript{$\ddag$} \\
   & Power-to-hydrogen & 24\textsuperscript{$\S\S$} & 0.44\textsuperscript{\cite{nep2019szenariorahmen}} & 0\textsuperscript{$\P$} & 0.95\textsuperscript{$\P$} & 1\textsuperscript{$*$} & 1.6 GW\textsuperscript{\cite{nep2019szenariorahmen}} & 200\textsuperscript{$\ddag$} & 100\textsuperscript{$\ddag$} & 150\textsuperscript{$\ddag$} \\
  E-mobility & E-mobility & 5\textsuperscript{\cite{pellinger2016merit}} & 0.07\textsuperscript{\cite{nitsch2012langfristszenarien,luca2014large}} & 0\textsuperscript{$*$} & 0.25\textsuperscript{$*$} & 1\textsuperscript{$*$} & 22 GW\textsuperscript{\cite{nep2019szenariorahmen}} & 84000\textsuperscript{$*$;\cite{pellinger2016merit}} & 16000\textsuperscript{$*$;\cite{pellinger2016merit}} & 10\textsuperscript{\cite{gils2015balancing}} \\
 \hline
 \end{tabular}
\end{table*}

\subsection{Load increase and decrease shares}\label{sec_met_s_inc_dec}
In case that there are no technical restrictions for decreasing or increasing the load for a specific appliance, $s_{dec}^c$ is set to zero and $s_{inc}^c$ is set to one. However, for some of the considered processes a complete shut-down due to temporary load shifting is not possible. In air separation plants for example this could damage the technical facilities or impair product quality~\cite{steurer2017analyse}. For such processes $s_{dec}^c$ is set to a value between zero and one.

For the industrial processes a revision outage of $5\%$ of the time of a year is assumed~\cite{steurer2017analyse}. This results in an average load increase limit of $s_{inc}^c=0.95$. As shown in Table~\ref{tab_dsm_parameters}, residential washing and drying appliances as well as domestic hot water heaters have very low utilisation rates. Due to usage preferences~\cite{darby2012social}, demand response can only lead to a limited increase of usage at a specific point of time, yielding $s_{inc}^c$ values below 0.2~\cite{steurer2017analyse}. The numerical values of $s_{dec}^c$ and $s_{inc}^c$ for all considered processes are provided in Table~\ref{tab_dsm_parameters}.

\subsection{Flexible share}\label{sec_flexible_share}
To account for socio-technical load shifting potential restrictions, the annual energy demand, $E_i^c$, was multiplied with the flexible share parameter, $s_{flex}^c$, in Eq.~\ref{eq_lamda} and Eq.~\ref{eq_l_i_t}.
The following restrictions are summarised in $s_{flex}^c$~\cite{steurer2017analyse}:
First, the organisational feasibility limits the potential, e.g., in the case that a change of working hours is necessary to allow for load shifting in the CTS or industrial sector. Second, the social acceptance is a limiting factor, for example if load shifting affects usage preferences of residential appliances. Third, the regulatory framework can hamper the implementation of load shifting. 
For each considered technology the $s_{flex}^c$ parameter values are presented in Table~\ref{tab_dsm_parameters}. 

\subsection{Load shifting costs}\label{sec_costs}
Load shifting costs can be divided into specific investment costs $c_{inv}$, annual fixed costs $c_{fix}$ and variable costs $c_{var}$~\cite{gils2016economic}. The investment is made up of the costs for information and communication technology (ICT) components, as well as installing and programming of the devices~\cite{steurer2017analyse}. The annual fixed costs are caused by maintenance works and the electricity consumption of the ICT components~\cite{steurer2017analyse}. The variable costs reflect compensations for losses in production outputs and comfort~\cite{gils2016economic}. In Table~\ref{tab_dsm_parameters} the assumed cost parameters and the used data sources are given.

We further analysed the load shifting investment cost for different PtH size classes in the residential sector, because the building structure significantly differs between rural and urban areas~\cite{heitkoetter2020regionalised}. 
As described in Section~\ref{sec_res_pth} we defined more than 700 residential building categories and assigned a heat demand value to each category. The buildings were grouped according to the heating types: single-storey heating, central heating and district heating. For the central heating technology, we distinguished between
three classes of installed thermal heating capacity: $\dot{Q}_{inst}<12.5~kW_{th}, 12.5~kW_{th}<\dot{Q}_{inst}<25~kW_{th}$ and $25~kW_{th}<\dot{Q}_{inst}$~\cite{heitkoetter2020regionalised}.
To determine the installed electric capacity, $P_{el}$, the heating capacity, $\dot{Q}_{inst}$, was divided by the annual average coefficient of performance of the electric heating, ${<}COP{>}$. Due to the greater energy efficiency~\cite{bloess2018power} and the higher number of new installations~\cite{heitkoetter2020regionalised}, we only considered heat pumps in this detailed cost investigation and neglected resistive heating technologies.  
The investment costs for the flexibilisation of heat pumps is estimated at $310~\text{\euro{}}$ in~\cite{pellinger2016merit}. This number is divided by the $P_{el}$ values of the different size classes to determine the specific investment costs, $c_{inv}$. The numerical values for $P_{el}$ and $c_{inv}$ are presented in Table~\ref{tab_hp_costs}.
\begin{table}[htbp]
\centering
\scriptsize
 \caption{Residential heat pumps investment costs for flexibilisation.}
 \label{tab_hp_costs}
 \begin{tabular}{l l l}
  \hline
  Heat pump size & $P_{el}[kW]$ & $c_{inv}[\text{\euro{}}/MW]$ \\
  \hline
  single-storey heating & 2.7 & 115000\\
  central heating small & 3.7 & 84000\\
  central heating medium & 5.5 & 57000\\
  central heating large & 14.2 & 22000\\
  \hline
 \end{tabular}
\end{table}


\subsection{Future scenario}\label{sec_met_fut_cen}
To model the future trend of the load shifting potential in Germany, we took into account a scenario for 2030. While the development of the installed capacities of the different technology classes was considered, the load profiles and load shifting parameters ($\Delta t$, $s_{util}$, $s_{dec}$, $s_{inc}$, $s_{flex}$), as well as the specific costs, were assumed not to change. Due to decarbonisation and the associated electrification of the heating and transport sector, the installed capacities of the e-Mobility, PtH and PtG technologies are expected to increase. We therefore used the German Grid Development Plan for 2030~\cite{nep2019szenariorahmen} as a reference for the future scenario, as it provides consistent predictions for the mentioned sector coupling technologies. Within the German Grid Development Plan there are different scenarios considered. In the present paper, we adopted scenario "B", which assumes a moderate speed of decarbonisation and flexibilisation of the energy system.

The other considered technology classes in the residential, CTS and industrial sector are very specific to load shifting and there are no detailed future projections for these technologies in the Grid Development Plan for 2030. We therefore used future projections from~\cite{steurer2017analyse} and~\cite{gils2015balancing} for the considered technologies. In these studies, the industrial production, specific demands, commercial electricity demand structure and residential appliances energy consumption were extrapolated until 2030, based on statistical data. 

For the model implementation of the future scenario, the regionalised maximum capacities of the technologies in 2018 were multiplied with the relative change until 2030, $x_{2030}$. For the e-mobility and PtG technologies there was only a very low installed capacity in 2018. We therefore did not apply the relative change factor, $x_{2030}$, but directly used the predicted capacity to be installed in 2030, $P_{2030}$. The numerical values are given in Table~\ref{tab_dsm_parameters}.

Concerning the residential, CTS and e-mobility technologies, we adopted the load regionalisation for the year 2018 also for the 2030 scenario. The regional distribution of such loads mainly depends on the population density, which we assumed not to significantly change until 2030. For the industrial load shifting technologies, we also assumed a constant regional distribution until 2030. Industrial sites at specific locations might shut down and new facilities might be opened in other regions of Germany. However, there were no reliable data available for predicting such relocation processes and therefore we neglected it.     

Regarding centralised PtH in district heating grids, as well as PtG plants, we used a different regionalisation methodology for the future scenario as for the status quo. As described in Section~\ref{sec_res_pth}, we used a plant specific regionalisation for the 2018 status quo, based on existing resistive heaters in district heating networks~\cite{christidis_2017_eneff}. For the 2030 scenario, we distributed the centralised PtH plants according to the overall heat demand of the district heating networks in the administrative districts~\cite{heitkoetter2020regionalised}. Also for the PtG plants the load regionalisation for the status quo is based on existing individual plants, which are research pilot projects in most cases~\cite{thema2019power}. To model the regional distribution in 2030, we used the industrial gas demand per administrative district~\cite{kunz2017reference}, because according to~\cite{nep2019szenariorahmen} PtG will mainly be used in the industrial sector.\footnote{For more information on the regionalisation of PtH in district heating and the PtG technologies in the future scenario refer to the supplementary material.}    

To model the influence of the ambient air temperature on the demand in the PtH sector, the weather year of 2018 was also used for the 2030 scenario.

\section{Results and Discussion}\label{sec_results}
\begin{figure*}[h!]
\subfloat[\label{sum_p_max_all_techs}Cumulated load increase potential $P_{max}$ for all technologies.]{\includegraphics[width=0.4\textwidth]{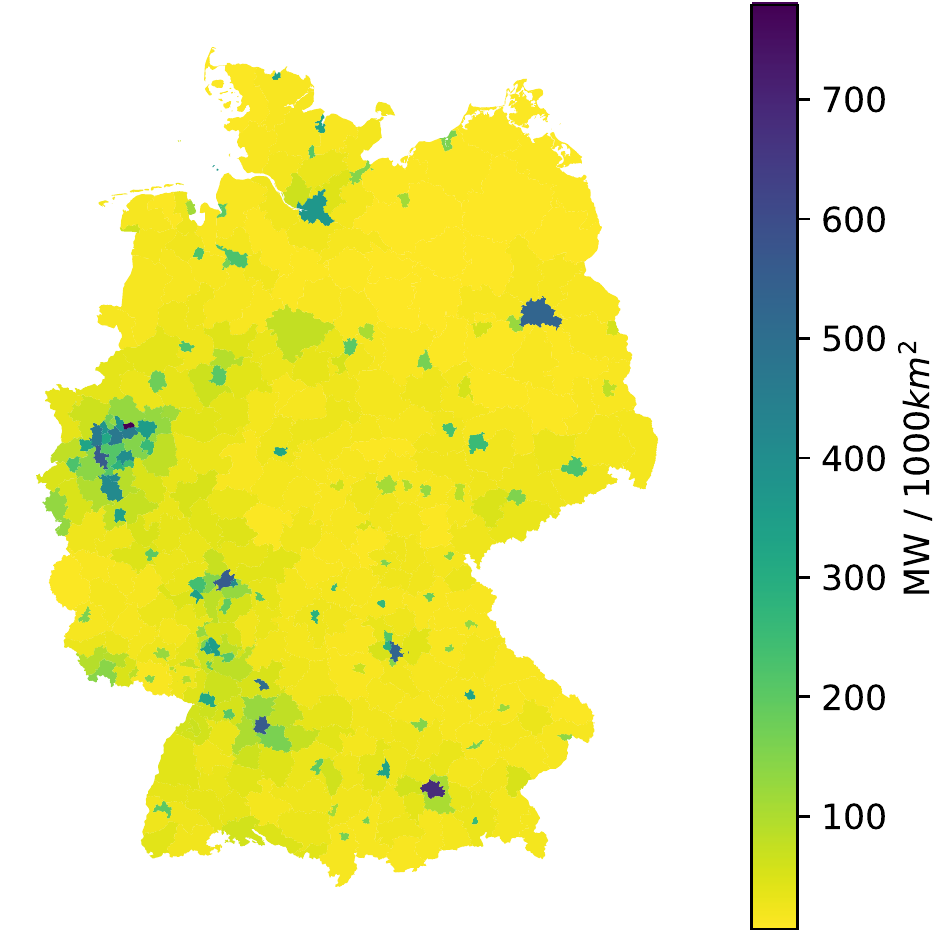}} 
\hspace*{\fill}
\subfloat[\label{sum_p_max_industry}Load increase potential $P_{max}$ for the industrial sector.]{\includegraphics[width=0.4\textwidth]{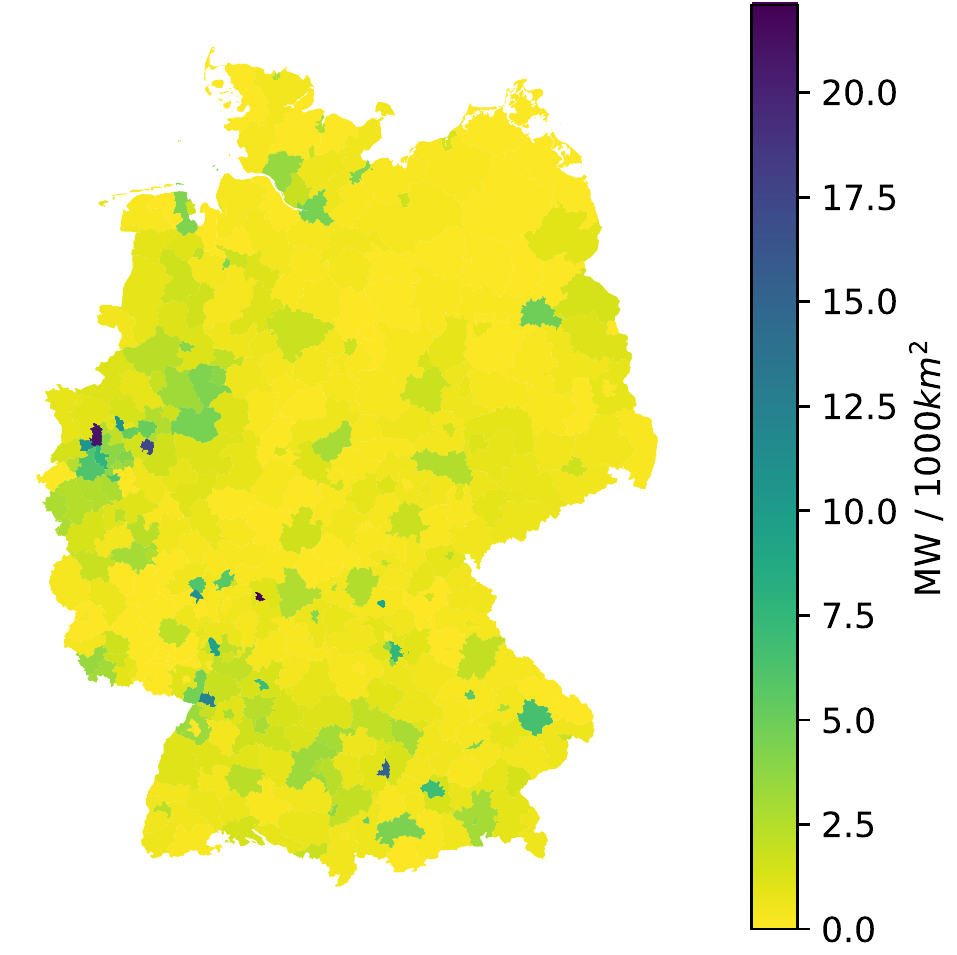}}
\caption{Geographical distribution of the load increase potential in the German administrative districts for the 2030 scenario.}
\end{figure*}
In this section we describe the results of the load shifting potential assessment. First, an overview of the spatial distribution of the potential is given and second, the temporal availability is assessed. Subsequently, regional cost-potential curves for the load shifting potentials are presented. Finally, the results of this paper are compared to the literature.  

\subsection{Spatial load shifting potential distribution}
The spatial distribution of the load shifting potential within the German administrative districts is analysed in the following paragraphs from a geographic perspective, as well as by looking at the frequency distribution. In Section~\ref{sec_comp_redis_curt} we compare the load shifting potential results with redispatch and curtailment key figures.

\subsubsection{Geographical potential distribution}
Figure~\ref{sum_p_max_all_techs} shows the geographical distribution of the 
cumulated maximum values for the load increase potential $P_{max}$ of all considered technologies in the German administrative districts for the 2030 scenario.\footnote{Refer to Section~\ref{sec_res_freq_dis} for a comparison of the 2018 and 2030 scenario.}
For better comparison each $P_{max}$ value was divided by the area of the administrative district. The results are given in $MW/1000~km^2$, as $1000~km^2$ corresponds approximately to the average area of the administrative districts in Germany of $891~km^2$.

The resulting spatial distribution reflects the population density with the metropolitan areas having the highest potential values. These are Berlin in the east of Germany, Hamburg in the north, the Ruhr area in the west and Rhein-Main area in the south-west. Furthermore, in Germany there are about 100 more major cities which form an independent administrative district. Most of these districts have a population of more than 100000 inhabitants, but a small territory~\cite{destatis2011census}, compared to the predominantly rural administrative districts. The resulting high area specific load shifting potential in the major city districts can be noted in Figure~\ref{sum_p_max_all_techs} by the small dark areas spread over the German territory.

Figure~\ref{sum_p_max_industry} shows the load increase potential distribution for only the industrial sector. Here, the metropolitan regions and other major cities do not dominate the potential as significantly as in the case of the cumulated potential of all sectors. Some of the predominantly rural districts show a relatively high industrial load increase potential. This can be noted by the administrative districts with a rather large territory having darker colours compared to Figure~\ref{sum_p_max_all_techs}. An explanation for this finding is that industrial sites, such as cement mills or paper plants are often located in rural areas~\cite{vdz2019zementwerke, vdp2017standorte}. 

For the residential, CTS, PtH and e-mobility demand sectors, no individual maps are shown here because the geographical distribution strongly correlates to the population density and thus corresponds to Figure~\ref{sum_p_max_all_techs}. For the status quo, the spatial distribution of the power-to-gas load increase potential is set by the locations of the power-to-gas plants in pilot projects~\cite{thema2019power}.

\subsubsection{Frequency distribution for administrative districts}\label{sec_res_freq_dis}
\begin{figure*}[h!]
\subfloat[\label{hist_sectors_p_max_av}Distribution of the potential values for load increase, $P_{max}$ (positive values), and load decrease, $P_{min}$ (negative values).]{\includegraphics[width=0.475\textwidth]{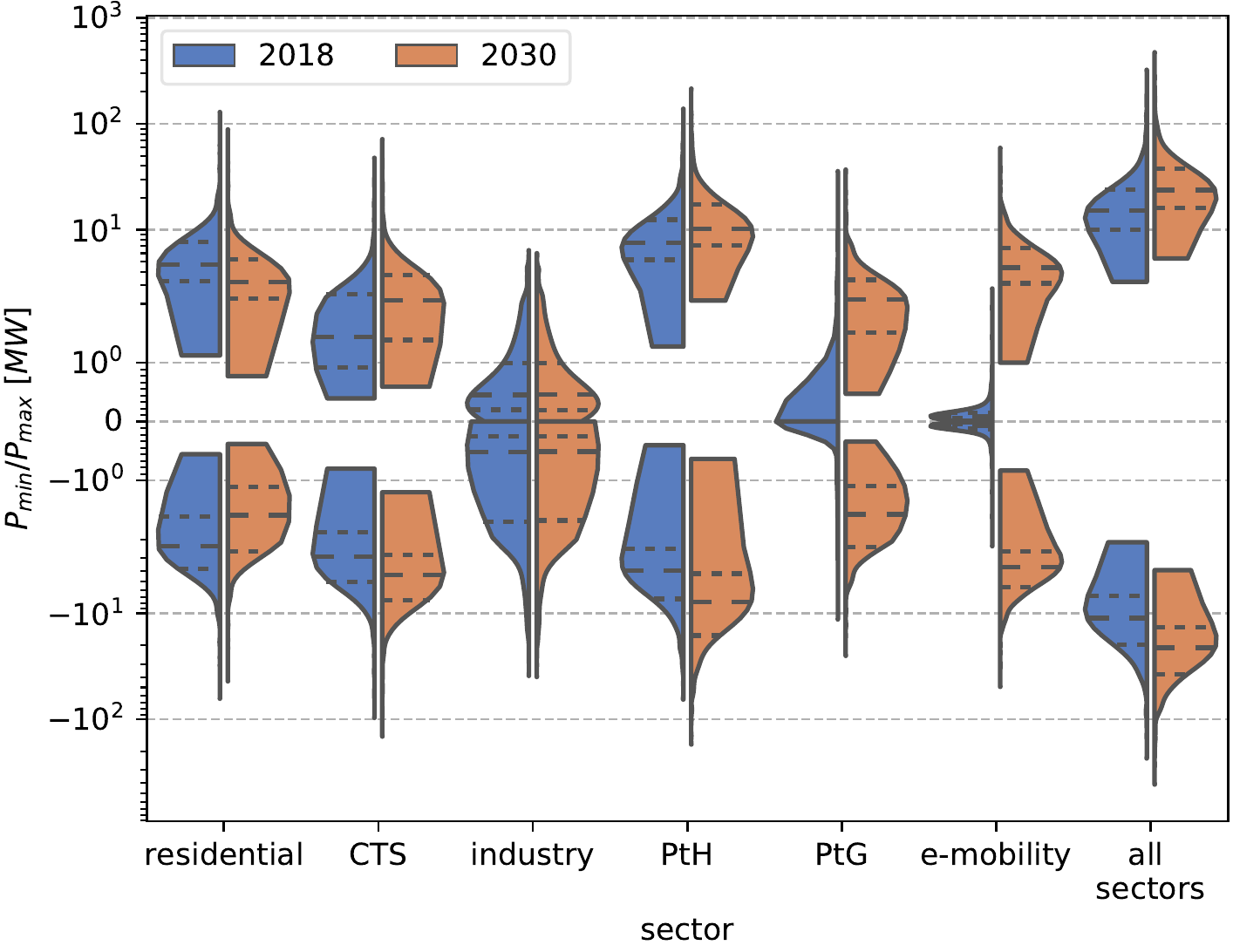}}
\hspace*{\fill}
\subfloat[\label{hist_sectors_e_max_av}Distribution of the potential values for energy preponing, $E_{max}$ (positive values), and energy postponing, $E_{min}$ (negative values).]{\includegraphics[width=0.475\textwidth]{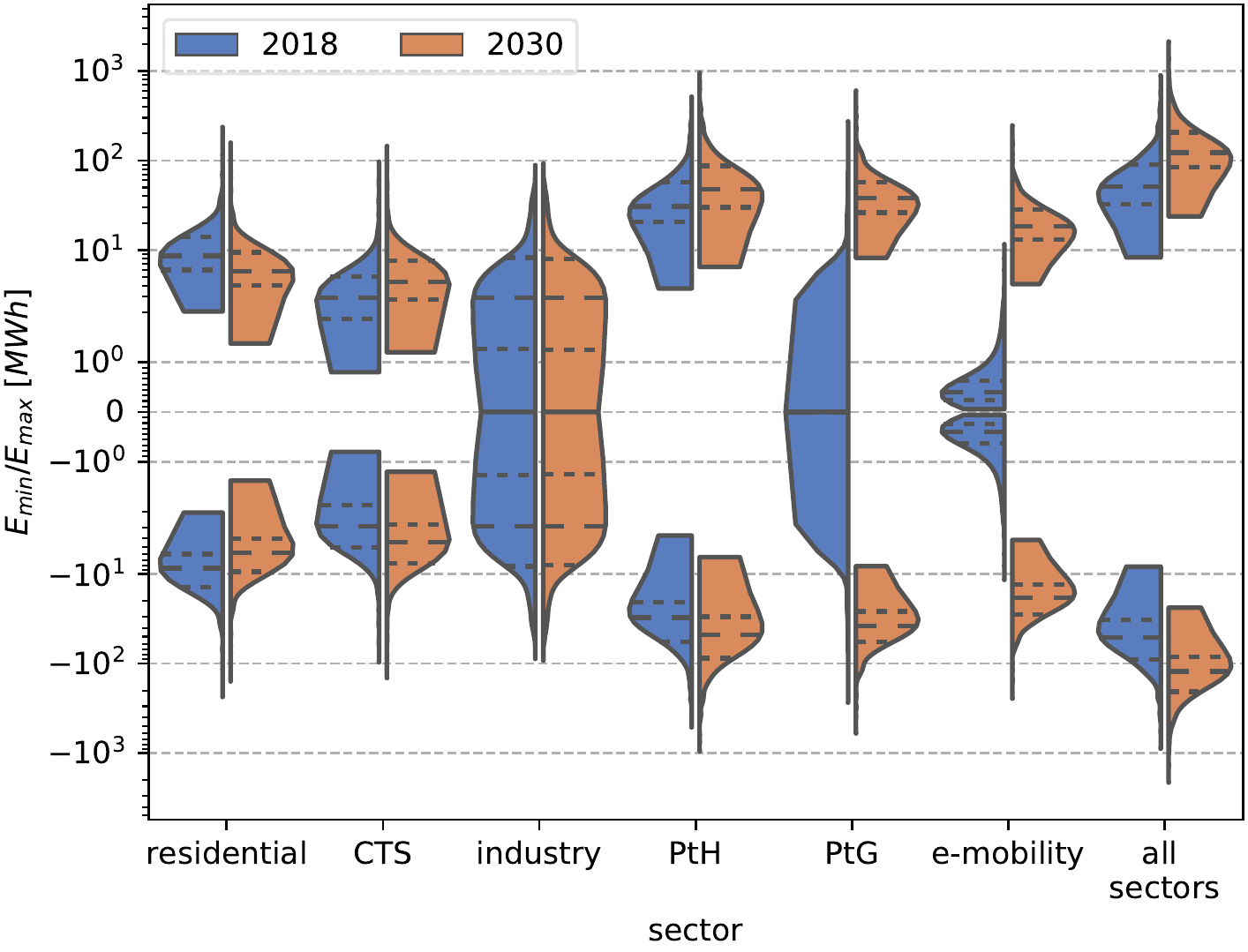}}
\caption{Distribution of the annual average load and energy shifting potential values over all German administrative districts in the different demand sectors for the status quo and 2030 scenario.}
\end{figure*}

The frequency distributions of the power and energy potential values for all German administrative districts are illustrated by violin plots in Figure~\ref{hist_sectors_p_max_av} and~\ref{hist_sectors_e_max_av}.
The distributions for load increase ($P_{max}$ and $E_{max}$) are plotted on the positive y-axis, the distributions for load decrease ($P_{min}$ and $E_{min}$) on the negative y-axis. The potential values of the individual technologies are aggregated sector wise, as described in Section~\ref{sec_methods}. The black dashed lines denote the quartiles of the distributions, where the second quartile corresponds to the median.

The distributions for all technologies show significant upper tails, which are caused by the largest major cities in Germany as, e.g., Berlin or Hamburg. The distribution shapes for the residential appliances, the CTS sector and the PtH technologies look relatively similar, as they are mainly driven by the number of inhabitants per district. In comparison, the distribution for the load shifting technologies in the industry sector shows a smaller upper tail, since industrial facilities are also often based in rural areas. The distribution for PtG shows the most significant upper tail, because there are only existing PtG plants in few administrative districts. 

For the year 2018 the PtH technologies provide the highest load increase potential, $P_{max}$, with a median value of $7.5~MW$ per administrative district, followed by residential appliances with $5~MW$ and the CTS sector with $1.2~MW$. While the potential is decreasing for residential appliances until 2030, due to improved energy efficiency, and stays approximately constant for the industrial sector, the potential for all other energy sectors is increasing. The strongest increase can be observed for e-mobility with a $P_{max}$ median value of $0.05~MW$ in 2018 and $4.5~MW$ in 2030. The median of the summed load increase potential for all sectors increases from $16~MW$ per administrative district in 2018 to $25~MW$ in 2030, the maximum $P_{max}$ value of all administrative districts increases from $220~MW$ to $390~MW$.

The load decrease potential, $P_{min}$, is higher than the load increase potential, $P_{max}$, for the CTS, industrial and PtG sector due to high utilisation rates. For the other sectors $P_{min}$ is lower than $P_{max}$ due to low utilisation rates. The median of the summed load decrease potential for all sectors increases from $-11~MW$ per administrative district in 2018 to $-21~MW$ in 2030, the maximum $P_{min}$ value of all administrative districts increases from $-250~MW$ to $-410~MW$.

The shapes of the distributions in Figure~\ref{hist_sectors_e_max_av} for the energy preponing and postponing potentials, $E_{max}$ and $E_{min}$, are similar as for the power buffers. Due to the higher time frame of management parameters, the potential values of the industrial, PtH, PtG and e-mobility sector are increased in comparison to the residential and CTS sector. The median of the summed load preponing potential, $E_{max}$, for all sectors increases from $50~MWh$ per administrative district in 2018 to $130~MWh$ in 2030. The maximum $E_{max}$ value of all administrative districts increases from $900~MWh$ to $2100~MWh$. The magnitudes of the $E_{min}$ values are equal to those of the $E_{max}$ values, as the same $\Delta t$ parameter is applied for both load preponing and postponing.

\subsubsection{Comparison with redispatch and curtailment key figures}\label{sec_comp_redis_curt}
As load shifting may be applied for avoiding grid congestion and curtailment of renewable energy sources~\cite{moura2010role}, we further assess our results with reference to these grid management measures. In the case of imminent grid congestion, one option is to apply redispatch. During this measure, the feed-in of a power plant on one side of the potentially overloaded grid element is decreased, while the feed-in of a power plant on the other side is increased~\cite{kamga2009regelzonenubergreifendes}. 

Under current regulations conventional power plants with a capacity larger than $10~MW$ take part in the redispatch process in Germany~\cite{hirth2019kosten}. Thus the calculated median value of the load increase potential per administrative district for all sectors of $16~MW$ in 2018 and $25~MW$ in 2030 exceeds the minimum power limit of $10~MW$. 
Only few administrative districts reach this threshold value using only a single load, for example the district of Heilbronn with a $100~MW$ resistive heater in a district heating grid~\cite{heitkoetter2020regionalised}.  
Another option would be to control smaller shiftable loads in an aggregated manner~\cite{paridari2015demand}, e.g., 2000 medium-scale heat pumps with an installed electric capacity of $5~kW$. Due to changes in the legal framework\footnote{%
Redispatch 2.0: Amendment of the Grid Expansion Acceleration Act (NABEG), 13th of May, 2019, \href{https://www.bgbl.de/xaver/bgbl/start.xav?start=\%2F\%2F*\%5B\%40attr_id\%3D\%27bgbl111043.pdf\%27\%5D\#__bgbl__\%2F\%2F*\%5B\%40attr_id\%3D\%27bgbl119s0706.pdf\%27\%5D__1592462510086}{https://www.bgbl.de}%
}, from 2021 on, power plants or storage units with a capacity larger than $100~kW$ will also be able to take part in the redispatch process. This threshold is exceeded by the load increase potential of all administrative districts in Germany, also when regarding single energy sectors. Only 20 medium-scale heat pumps would need to be aggregated to reach the $100~kW$ threshold.

German transmission grid operators are obliged to publish each individual redispatch measure online.\footnote{\href{https://www.netztransparenz.de/EnWG/Redispatch}{https://www.netztransparenz.de/EnWG/Redispatch}} We averaged the power and energy of all individual redispatch measures between 2013 and 2020, yielding an average power of $230~MW$ and an average energy of $1700~MWh$. 
Approximately 9 administrative districts having a load increase potential of $25~MW$ (median value for the 2030 scenario) would need to be aggregated to provide a power of $230~MW$. To reach the average redispatched energy of $1700~MWh$, the potential of 13 administrative districts with a potential of $130~MWh$ would need to be aggregated. Consequently the average load shifting potential of single administrative districts seems to be to small to fully cover average redispatch measures.

However, load shifting may play an auxiliary role for avoiding redispatch of power plants and curtailment of renewable energy sources. As an example, we consider classical $3~MW$ wind turbines~\cite{hirth2016system} to be curtailed from full load to zero load, due to grid congestion. When load increase is used to utilise the excess feed-in from wind power, the load increase potential of an administrative district of $25~MW$ would suffice to avoid the curtailment of 8 wind turbines. The median energy shifting potential of $130~MWh$ per administrative district would suffice to avoid the curtailment of 8 wind turbines for 5 hours.  

\begin{figure*}[h!]
\centering\includegraphics[width=\linewidth]{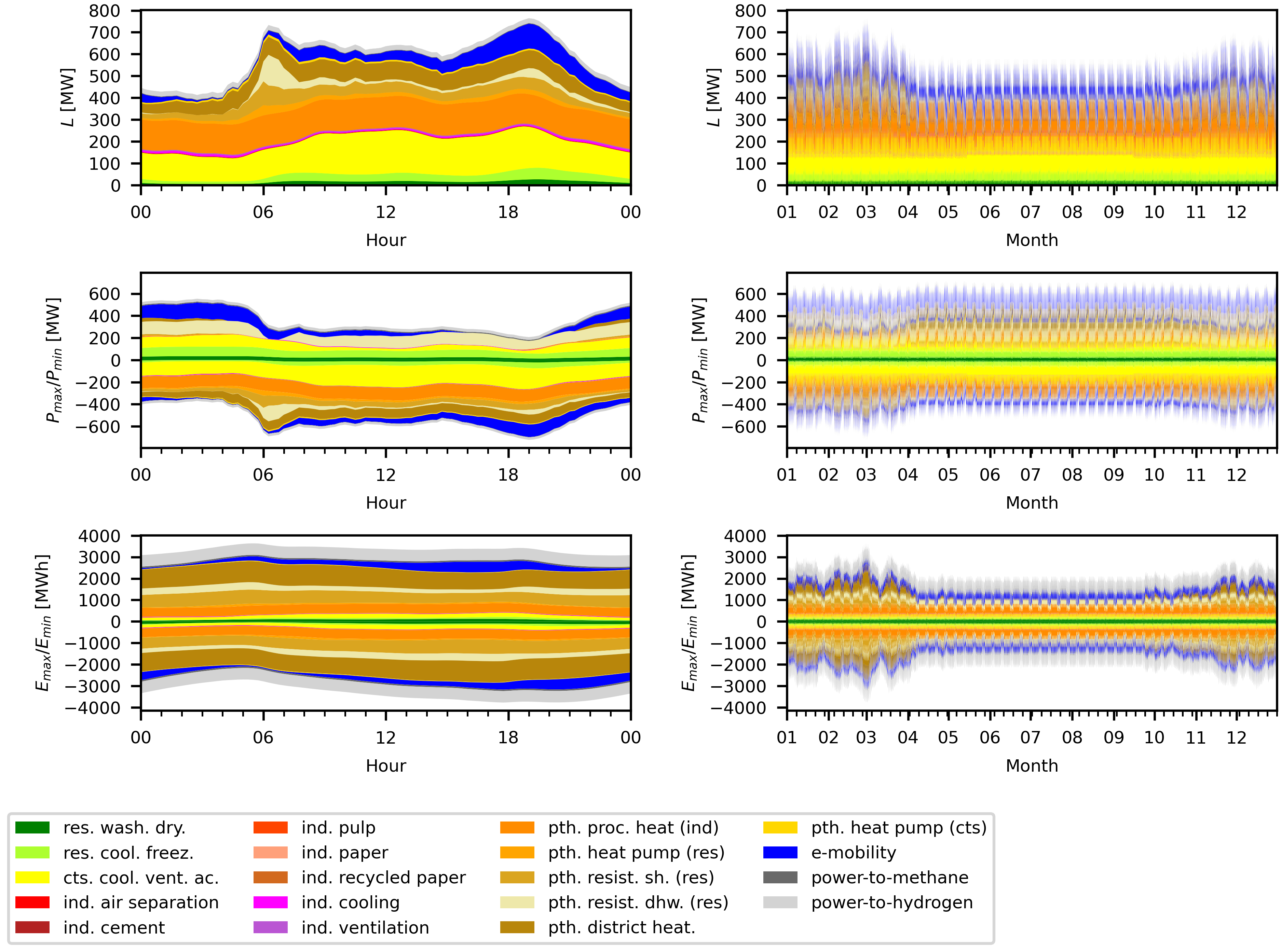}
\caption{Temporal distribution of the scheduled load (top row) for the city of Berlin, load increase and decrease potential (middle row) and energy preponing and postponing potential (bottom row); Left column: cold winter day (1\textsuperscript{st} of March, 2018); Right column: Yearly time series.}
\label{temporal_distribution}
\end{figure*}

\subsection{Temporal load shifting potential distribution}
We analyse the temporal distribution of the load shifting potential using the example of the city of Berlin for the 2030 scenario.
The top row of Figure~\ref{temporal_distribution} shows the time series of the scheduled load, $L$, the middle row shows the load increase and decrease potentials, $P_{min}$ and $P_{max}$, and the bottom row shows the energy preponing and postponing potentials, $E_{min}$ and $E_{max}$.
The left column illustrates the intraday profiles for an exemplary cold winter working day (weather data of 1\textsuperscript{st} of March, 2018).
While the scheduled load of most of the considered technologies is relatively constant, 
the residential PtH technologies and e-mobility show significant load peaks in the morning as well as evening hours.\footnote{For a discussion of the peaks in the PtH load profiles refer to Section~\ref{sec_crit_app}.} 
This leads to high load decrease potential during these time periods, e.g., a $P_{min}$ value of $-700~MW$ at 06:00, and a low load increase potential, e.g., a $P_{max}$ value of $300~MW$ at 06:00. Conversely, the low scheduled load during night time leads to a high load increase and a low load decrease potential.

As defined in Eq.~\ref{eq_e_max} and~\ref{eq_e_min}, for the calculation of $E_{max}$ and $E_{min}$, the scheduled load is integrated over time, using the time frame of management, $\Delta t$ as integration limit. This reduces the influence of short term load peaks and results in smoother $E_{max}$ and $E_{min}$ profiles in comparison to the profiles of $P_{max}$ and $P_{min}$. However, there is a slight $E_{max}$ peak of $3500~MWh$ at 05:30 in the morning and an $E_{min}$ peak of $-3500~MWh$ at 17:30 in the afternoon. This can be explained by $\Delta t$ being $\leq 12~h$ for most technologies and the scheduled load being higher during day time than during night time. Thus, most energy can be preponed from day to the morning and postponed from day to the evening.  

The right column of Figure~\ref{temporal_distribution} shows the load shifting potential distribution during the course of the year.
Due to the changing space heating demand, PtH technologies are more utilised in winter and less utilised in summer. 
Considering the summed potential for all technologies, this leads to an average load decrease value, $P_{min}$, that is approx. $25\%$ higher on a cold winter day than on a summer day. In contrast, the load increase potential $P_{max}$ is approx. $25\%$ higher on a summer day than on a cold winter day.

The values of $E_{max}$ and $E_{min}$ for the PtH technologies are both increased during winter, because the higher utilisation leads to a higher amount of energy that can be preponed and postponed. The summed potential for all technologies is twice as high on a cold winter day than on a summer day. This significant difference between summer and winter is due to the relatively high share of PtH technologies in the $E_{max/min}$ potential and is caused by the large $\Delta t$ value for PtH, compared to the other technologies.

\begin{figure*}[h!]
\centering\includegraphics[width=\linewidth]{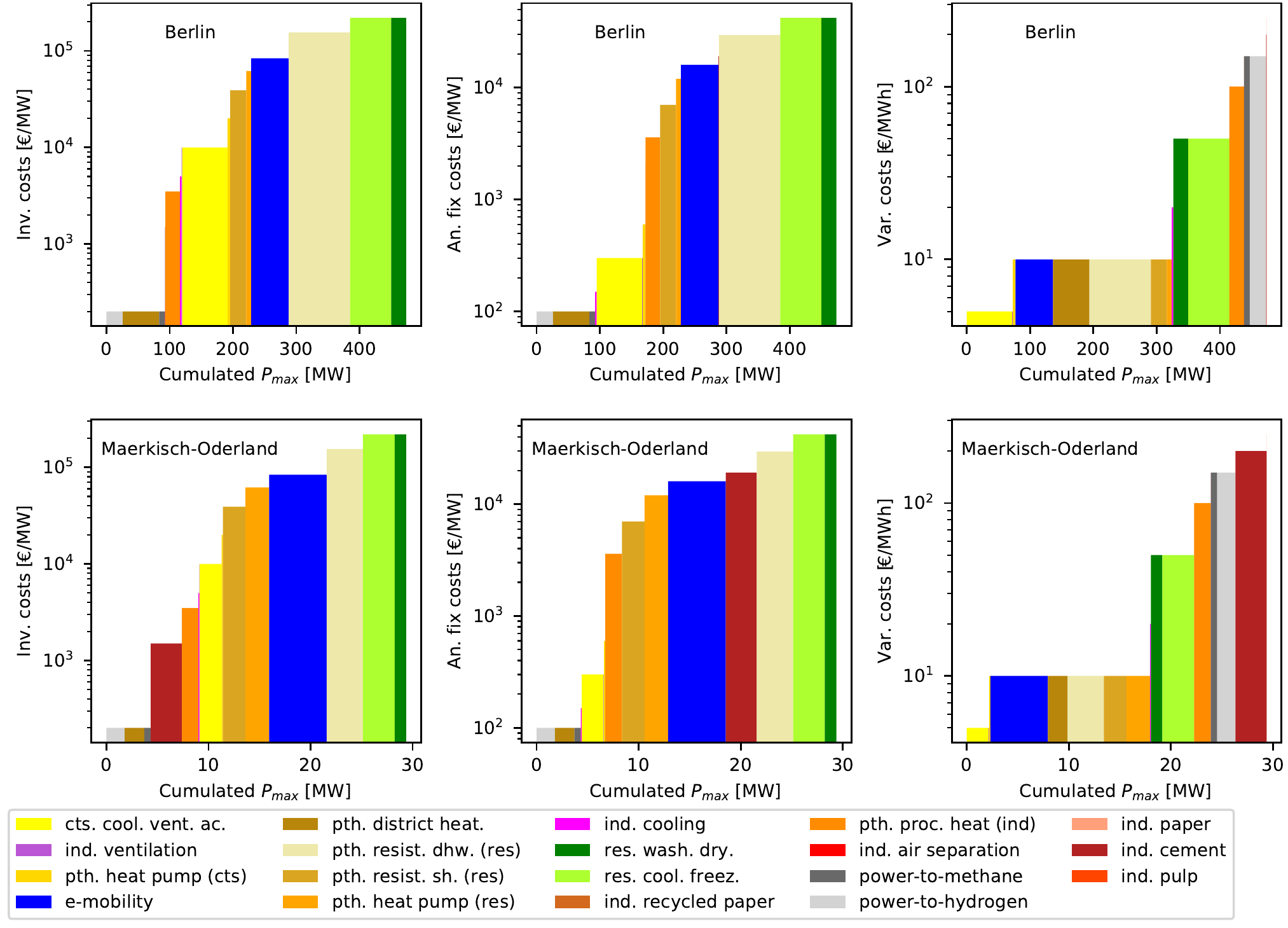}
\caption{Load shifting cost-potential curves for the city of Berlin (top row) and the rural district Maerkisch-Oderland (bottom row) for the 2030 scenario; Left column: Investment costs; Middle collumn: Fixed costs; Right column: Variable costs.}
\label{merit_order_all_tech}
\end{figure*}

\subsection{Regional cost-potential curves}
As described in Section~\ref{sec_costs}, for each considered administrative district, cost-potential curves for load shifting were calculated. 
Figure~\ref{merit_order_all_tech} shows the investment cost, annual fixed cost and variable costs over the cumulated load increase potential $P_{max}$. In order to compare regions with different population density, we present the results for Berlin ($4000~residents/km^2$) and the neighbouring rural district Maerkisch-Oderland ($90~residents/km^2$) for the 2030 scenario as an example.

In general, industrial processes and PtH in district heating have the lowest specific investment costs due to the high installed electric capacity per facility, whereas residential appliances have low installed capacities and high specific investment costs (see also Table~\ref{tab_dsm_parameters} for the numerical values). On the other hand, industrial processes have high variable costs due to impairment of production processes~\cite{steurer2017analyse}, while ventilation and heating appliances have low variable cost, because of less user interference.

While the resulting overall load shifting potential for Berlin is $475~MW$, it is only $29~MW$ in Maerkisch-Oderland, due to the lower population density. In general, the distribution of the potential for the different technologies is similar for Berlin and Maerkisch-Oderland. However, the most striking difference between the two regions is the share of the industrial processes' potential. As there is a large scale cement production facility in the Maerkisch-Oderland district~\cite{vdz2019zementwerke}, but the overall load shifting potential is small, the industrial processes' load shifting potential share is $11\%$. In Berlin, the share of industrial processes in the overall load shifting potential is only $1\%$.    

In order to assess the costs for load shifting, we use the costs for the curtailment of renewable energy sources as a reference. According to the German Federal Network Agency~\cite{bnetza2020quartalsbericht}, $6482~GWh$ of electricity from renewable energy sources were curtailed in 2019, leading to compensation payments of $709.5~million~\text{\euro{}}$. Dividing the compensation payments by the curtailed energy yields average costs of $110~\text{\euro{}}/MWh$. 
We further assume that load shifting for avoiding curtailment of renewable energy sources is only used if the variable costs are lower than the average curtailment compensation costs.
All considered load shifting technologies meet this condition, except air separation, cement milling, wood pulping, paper production and PtG.
In this case, the load shifting potential for the city of Berlin would be reduced by $7\%$, which is mainly caused by the PtG technology. For the Maerkisch-Oderland district, $19\%$ of the load shifting potential would be excluded due to the share of the cement and PtG plants.

To further analyse regional influences on the load shifting cost-potential curves, we examined the heat pump technology more closely.
\begin{figure}[h!]
\centering\includegraphics[width=\linewidth]{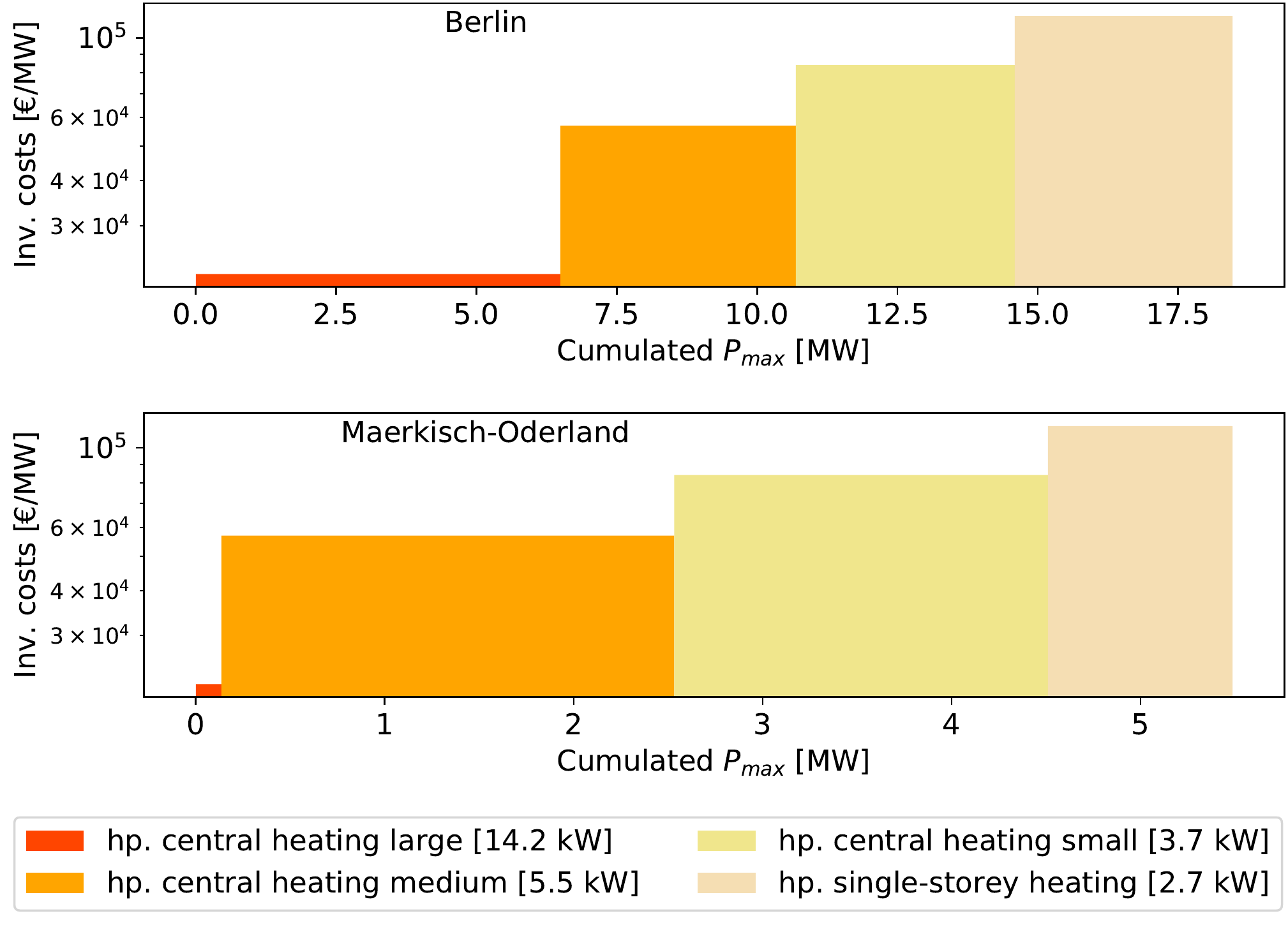}
\caption{Investment costs for the flexibilisation of residential heat pumps of different size classes.}
\label{merit_order_hp_size_classes}
\end{figure}
As described in Section~\ref{sec_costs}, we introduced four size classes of heat pumps and assigned specific investment costs for flexibilisation for each class. Figure~\ref{merit_order_hp_size_classes} shows the resulting specific investment costs over the cumulated load increase potential $P_{max}$ for Berlin and the Maerkisch-Oderland district.
While in Berlin there is a high number of multi-family houses, rural areas, such as Maerkisch-Oderland are dominated by single-family houses~\cite{heitkoetter2020regionalised}. This leads to a $35\%$ share of large central heatings in Berlin with low specific investment costs for heat pump flexibilisation of $22000~\text{\euro{}}/MW$. 
On the contrary, in Maerkisch-Oderland there is only a $2.5\%$ share of large-scale heat pumps and $44\%$ of heat pumps in medium central heatings, with higher specific investment costs of $57000~\text{\euro{}}/MW$~\cite{heitkoetter2020regionalised}. Next, we calculated weighted cost averages, using the $P_{max}$ capacities for each heat pump size class as weight. This results in average cost of $63~\text{\euro{}}/MW$ for Berlin and $23\%$ higher costs of $76~\text{\euro{}}/MW$ for Maerkisch-Oderland.

\subsection{Critical appraisal}\label{sec_crit_app}
\begin{figure}[h!]
\centering\includegraphics[width=\linewidth]{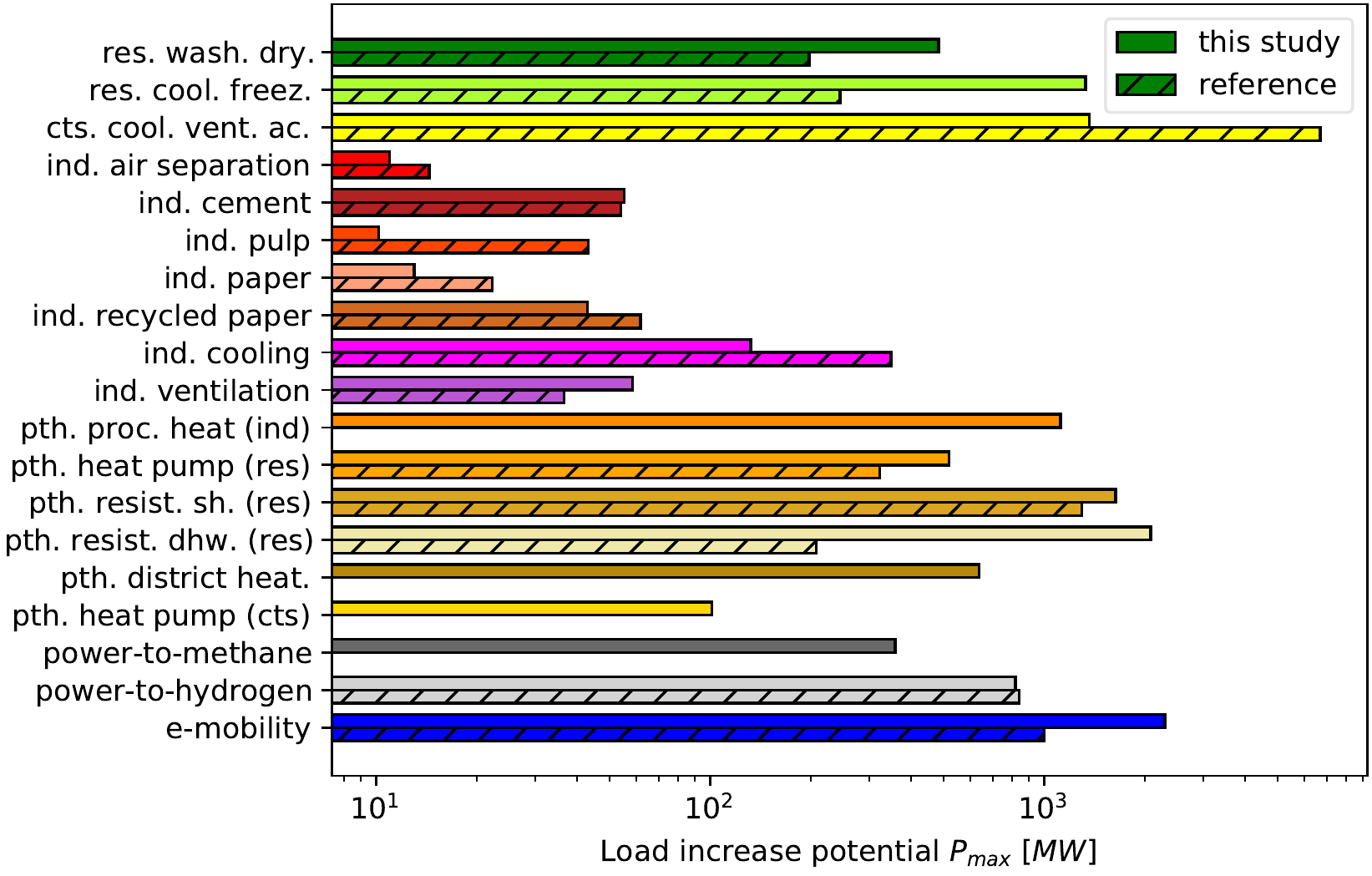}
\caption{Validation of the results for the annual average load increase potential values, $P_{max}$, in all of Germany for the 2030 scenario with literature reference values; The following validation references were used for the considered technologies: 
res. wash. dry.~\cite{steurer2017analyse},
res. cool. freez.~\cite{steurer2017analyse},
cts. cool. vent. ac.~\cite{gils2015balancing},
ind. air separation~\cite{steurer2017analyse},
ind. cement~\cite{steurer2017analyse},
ind. pulp~\cite{steurer2017analyse},
ind. paper~\cite{steurer2017analyse},
ind. recycled paper~\cite{steurer2017analyse},
ind. cooling~\cite{steurer2017analyse},
ind. ventilation~\cite{gils2015balancing},
pth. heat pump (res)~\cite{steurer2017analyse},
pth. resist. sh. (res)~\cite{steurer2017analyse},
pth. resist. dhw. (res)~\cite{steurer2017analyse},
power-to-hydrogen~\cite{pellinger2016merit},
e-mobility~\cite{pellinger2016merit}.
}
\label{validation_p_max}
\end{figure}

To validate the results of this paper, we compared the calculated values for the load increase potential, $P_{max}$, with literature reference values~\cite{gils2015balancing,steurer2017analyse,pellinger2016merit}, as depicted in Figure~\ref{validation_p_max}. For several considered technologies the resulting values of this paper are higher than those in the literature and for other technologies the values are lower, thus there is no general under- or overestimation.
Concerning the residential appliances, the results of this study are $300~MW$ higher for washing and drying machines and $1000~MW$ higher for fridges and freezers than in~\cite{steurer2017analyse}. This can be explained by the different values assumed for $s_{util}$ and $s_{inc}$. For washing machines $s_{util}$ and $s_{inc}$ have very low values ($<5\%$) and $P_{max}$ is approx. proportional to $s_{inc} - s_{util}$. Therefore differences of only several percent in the input parameters lead to high deviations in the results for $P_{max}$.

For the industrial processes the differences are between $5-120~MW$ for all considered technologies and are probably due to deviations in the estimated installed capacity of the processes.
Regarding the PtH technologies, the results of this paper and~\cite{steurer2017analyse} differ by $200-1800~MW$. The largest difference is present for resistive DHW heating. A reason for the deviation is that in this paper it was assumed that the $25\%$ share of resistive DHW heaters which is equipped with a thermal storage~\cite{apel2012notwendiger} is also suitable for load shifting, while in \cite{steurer2017analyse} only a $12\%$ flexible share is assumed. Furthermore, we applied a utilisation rate of $3\%$~\cite{gils2015balancing} and an hourly profile~\cite{heitkoetter2020regionalised}, while \citet{steurer2017analyse} differentiated between two phases of the day with a constant load profile each and differing utilisation rates.

For power-to-hydrogen and e-mobility no values are given in~\cite{gils2015balancing} and~\cite{steurer2017analyse} and we therefore used reference values from~\citet{pellinger2016merit}. In~\cite{pellinger2016merit} only the installed capacity of power-to-hydrogen plants is given, no value for the load increase potential. 
We therefore assumed a constant load profile and a utilisation rate of $44\%$, which yields a load increase potential of approx. $800~MW$ and is equal to the result of this study. 
The calculated load increase potential for e-mobility is about two times higher in this paper compared to~\cite{pellinger2016merit}. The reason for this is that we assumed an electric vehicle fleet of $6~million$ cars for 2030~\cite{nep2019szenariorahmen} and~\citet{pellinger2016merit} estimated $3~million$ electric vehicles to be existing. No literature reference values were found for load increase of PtH facilities covering industrial process heat or district heating demand, as well as for heat pumps in the CTS sector.   
Since the results for $P_{min}$, $E_{max}$ and $E_{min}$ mainly depend on the same input parameters as $P_{max}$, we did not carry out a separate validation for these potentials.  

Regarding the temporal distribution of the load shifting potential, there are significant peaks in the load increase and decrease time series for the residential space heating and DHW power-to-heat technologies. The reason for this is that the utilised daily load profiles~\cite{heitkoetter2020regionalised} are based on measurements of the heat flow and return at the heat exchangers and represent the actual demand of the residents. 
Since the considered electric heating devices are equipped with thermal storage~\cite{henze2004evaluation}, the electricity consumption can be decoupled from the actual demand.
Thus, electric heating devices often run at rather constant load and increased demand during night time, to support the operation of conventional power plants, which is incentivised by tariff design~\cite{darby2018smart}.  
Such a load profile might be used as reference scheduled load for applying additional load shifting, e.g. for avoiding curtailment of renewable energy sources. This would result in smoother profiles for the load increase and decrease potentials. However, we did not use such a load profile as a reference, as it already includes a preceding load shifting, compared to the actual demand of the residents.      

Furthermore, the utilised input parameters given in Table~\ref{tab_dsm_parameters} for modelling demand response and the associated costs are fraught with uncertainty. As load shifting is not widely used currently, the parameters need to be determined by small-scale pilot projects and surveys~\cite{steurer2017analyse}. E.g. for heat pumps, \citet{steurer2017analyse} estimates the uncertainty for the time frame of management parameter, $\Delta t$, at $\pm 20\%$. For several technologies, as for example PtH in district heating, industrial process heat and PtG, not all load shifting parameters are given in the literature. Therefore parameters of comparable technologies were adopted, which may cause inaccuracies in the determined load potential values.

\section{Conclusion and Outlook}\label{sec_conclusion_outlook}
One option to avoid the curtailment of renewable energies is to cover excess feed-in by load shifting.
As curtailment is often a local phenomenon, in this work we determined the regional load shifting potential and the associated costs for the 401 German administrative districts, considering 19 suitable technologies.
The results are provided with a temporal resolution of 15 minutes and a status quo analysis for 2018, as well as a 2030 scenario are considered.

In contrast to other studies in this field, all data and the developed source code are published open source. We highlighted the load shifting potential of sector coupling technologies by taking into account power-to-heat, power-to-gas and e-mobility in addition to the considered conventional loads. Further, we put a special focus on the regionalisation of the residential building stock, considering more than 700 building types.

The highest load shifting potential is provided by the power-to-heat technologies, the lowest by the considered industrial processes suitable for load shifting. 
The strongest growth of the potential from 2018 to 2030 can be observed for power-to-gas and e-mobility.
The spatial potential distribution is mostly governed by the population density. For industrial processes, there are also relatively high potential values in rural areas.
For the 2030 scenario, the load increase potential values range from $5 - 470~MW$ per administrative district and the median value is $25~MW$. 
In comparison, the lower power threshold for power plants to take part in redispatch is $10~MW$ and the average power of redispatch measures in the German transmission grid is $230~MW$. Load shifting is thus not a real alternative to redispatch of power plants, but may play an auxiliary role. When compared to curtailment, the median load increase potential value per administrative district would suffice to avoid the curtailment of 8 wind turbines with 3 MW installed capacity.   


The temporal distribution of the load increase and decrease potentials shows significant peaks in the morning and afternoon hours, caused by the demand time series of PtH and e-mobility. Due to the changing space heating demand in the course of the year, load decrease potentials are approx. $25\%$ higher on a cold winter day than on a summer day.

Industrial processes, power-to-gas and power-to-heat in district heating have the lowest load shifting investment cost, due to the largest installed capacities per facility. Ventilation, cooling and heating appliances have the lowest variable costs due to the least user interference. 
Further distinguishing between different size classes of the installed capacity of heat pumps leads
to $23\%$ lower average investment costs for heat pump flexibilisation in the city of Berlin compared to the rural district of Maerkisch-Oderland.
The variable costs of most considered load shifting technologies remain under the average compensation costs for curtailment of renewable energies of $110~\text{\euro{}}/MWh$.

As the load shifting potentials time series determined in this study are published open source, they can be used by other researchers as boundary conditions for load shifting dispatch in energy system models.
The provided investment, fixed and variable costs for load shifting can be further used for an economic assessment considering the full life time of the technologies.  
For industrial process heat, power-to-gas and power-to-heat in district heating, detailed technical studies are required for deriving more reliable load shifting parameter values.

\section*{Acknowledgements}
The first author gratefully acknowledges the financial support provided by the Foundation of German Business (sdw) through a PhD scholarship. 
The work of the second and third author was carried out as part of the enera project, which is funded by the Federal Ministry of Economic Affairs and Energy (BMWi, grant no. 03SIN317). 
Further, the authors thank Hans Christian Gils, Karl-Kien Cao and Jan Jebens for the fruitful discussions on regionalisation and load shifting, as well as Niklas Wulff for valuable hints regarding e-mobility modelling.

\appendix
\section{Supplementary Material}
Supplementary data associated with this article can be found at: \href{https://doi.org/10.5281/zenodo.3988921}{https://doi.org/10.5281/zenodo.3988921}.
\bibliographystyle{model1-num-names}
\bibliography{main.bib}

\end{document}